\documentclass[preprint,12pt,showpacs,aps,floatfix,superscriptaddress]{elsarticle}
\usepackage{amsmath,amssymb,graphicx,bm,color,mathrsfs,verbatim}
\pdfoutput=1

\newcommand{\be}{\begin{equation}}
\newcommand{\ee}{\end{equation}}

\begin{document}

\begin{frontmatter}

  \title{Emergent Newtonian dynamics and the geometric origin of mass}
  \author[PSU,BU]{Luca D'Alessio\corref{cor1}}
  \author[BU]{Anatoli Polkovnikov\corref{cor2}}
  \address[PSU]{Department of Physics, The Pennsylvania State University, 
University Park, PA 16802, USA}
  \address[BU]{Physics Department, Boston University, Boston, MA 02215, USA}
  \cortext[cor1]{E-mail address: dalessio@bu.edu}
  \cortext[cor2]{E-mail address: asp@bu.edu}

\begin{abstract}
We consider a set of macroscopic (classical) degrees of freedom coupled to an 
arbitrary many-particle Hamiltonian system, quantum or classical. These degrees 
of freedom can represent positions of objects in space, their angles, shape 
distortions, magnetization, currents and so on. Expanding their dynamics near 
the adiabatic limit we find the emergent Newton's second law (force is equal to 
the mass times acceleration) with an extra dissipative term. In systems with 
broken time reversal symmetry there is an additional Coriolis type force 
proportional to the Berry curvature. We give the microscopic definition of the 
mass tensor relating it to the non-equal time correlation functions in 
equilibrium or alternatively expressing it through dressing by virtual excitations in the system. In the classical (high-temperature) limit the mass tensor is given 
by the product of the inverse temperature and the Fubini-Study metric tensor 
determining the natural distance between the eigenstates of the Hamiltonian. For 
free particles this result reduces to the conventional definition of mass. This 
finding shows that any mass, at least in the classical limit, emerges from the 
distortions of the Hilbert space highlighting deep connections between  any 
motion (not necessarily in space) and geometry. We illustrate our findings with 
four simple examples.
\end{abstract}


\end{frontmatter}
\tableofcontents

\section{Introduction}
Newton's second law, $F= m \ddot X$, applies to a wide range of phenomena and it 
is the cornerstone of the classical physics.  This equation certainly describes 
the motion of particles in free space but it also predicts the dynamics of 
macroscopic objects like rotating bodies where $X$ represents the angle and $m$ 
is the moment of inertia or the behavior of the electrical current in a LC 
circuits where $X$ is a current and $m$ is a combination of capacitance and 
inductance.

However, as we all know, when an object moves through a medium it dissipates 
energy into the surrounding environment increasing its entropy and, as a result, 
it slows down and eventually stops. 
In this case, to properly describe the object's dynamics, Newton's law need to 
be supplemented with a dissipative (drag) term:
\begin{equation}
m \ddot X +\eta \dot X= F.
\label{eqNewton}
\end{equation}
Drag is not specific to the motion in real space and it is present every time a 
physical quantity changes in time. For example, when the magnetic flux through a 
coil is increased in time, the environment (electrons in the coil) will react by 
producing a drag force which opposes this change (i.e. the Faraday’s law). 
Likewise coherent spin oscillations always decay in time due to dissipation in 
the spin environment and so on.

A pragmatic and very successful approach consists in seeing the mass $m$ and the 
dissipation $\eta$ in Eq.~\eqref{eqNewton} as
emergent properties (i.e. phenomenological parameters) which need to be fitted 
to reproduce the correct dynamics of macroscopic degrees of freedom. This 
approach is conceptually unsatisfactory and, for this reason, a lot of research 
effort has been focused on trying to
 {\it derive} the mass and the dissipation from a more fundamental theory (see 
e.g. Refs.~\cite{wilczek_2004,berry_robbins,JAR,avron_2011}).
Normally one starts from the microscopic approach for both the object and the 
medium and using various approximations derives effective equations of motion 
for the macroscopic degree of freedom (d.o.f.)~\cite{Feynman_1954, 
Caldeira-Legget,Lindblad,Markov,Petruccione}.  From this approach one finds both 
renormalization of the bare mass by dressing with elementary excitations and the 
drag force (see for example \cite{Kamenev}). For example, in the Landau-Fermi 
Liquid theory \cite{Fermi-Liquid1,Fermi-Liquid2} the electron-electron 
interaction lead to a renormalization of the bare electron mass (and other 
dynamical properties).  

\begin{figure}
\includegraphics[width=0.90\columnwidth]{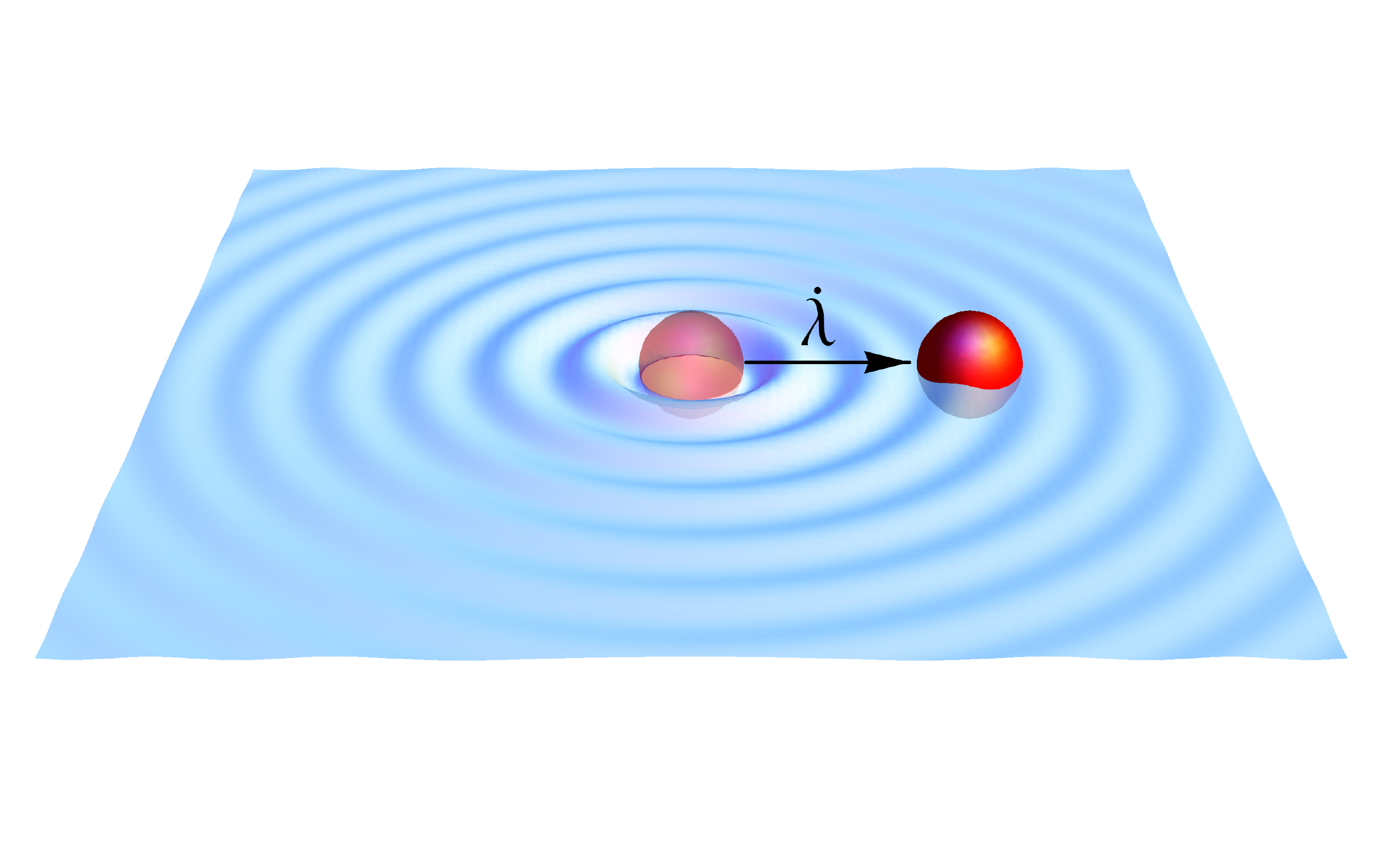}
\caption{(Color on-line) Schematic setup considered. A macroscopic d.o.f. is 
initially at equilibrium (light red ball) with a complex environment (indicated 
by the wave-background). Then it starts moving with velocity $\dot \lambda$ 
through the environment, which has to rearrange, and generates both a drag force 
and a renormalization of the mass of the d.o.f.. The drag coefficient is related 
to real transitions between eigenstates of the environment (see 
Eq.~\eqref{eq:main-eta}) while the mass is related to virtual transitions which 
describe dressing of the d.o.f. by the virtual exvitations of the environment (see Eq.~\eqref{main_kappa}).}
\label{cartoon}
\end{figure}

In this work we take a different approach (see \cite{berry_robbins} and 
references therein), where we use statistical equilibrium of a complex system as 
a starting point. Namely we consider an arbitrary many-body system with possibly 
infinitely many degrees of freedom interacting with few macroscopic parameters 
$\vec\lambda(t)$, which are allowed to slowly change in time (see 
Fig.\ref{cartoon}). We assume that if $\vec\lambda$ is constant the system 
equilibrates at some temperature (as we will discuss later this assumption can 
be further relaxed). By extending the Kubo linear response theory~\cite{Kubo} to 
such setups we provide a framework to compute the energy exchange between the 
many-body system and macroscopic d.o.f. in time derivatives of $\vec\lambda(t)$. 
In the first two orders (for systems with time-reversal invariance) the 
resulting expression takes the form of the Newton's second law with the 
additional drag term:
\begin{equation}
\kappa_{\nu\mu} \ddot \lambda_\mu+\eta_{\nu\mu} \dot\lambda_\mu =M_\nu,
\label{emergent}
\end{equation}
where $\kappa_{\nu\mu}$ is the mass tensor, $\eta_{\nu\mu}$ is the drag tensor, 
and $M_\nu$ is the force (we choose this notation to reserve the letter $F$ for 
the Berry curvature). 
In the absence of symmetries all three coefficients can explicitly depend on 
$\vec\lambda$ predicting nontrivial dynamics. We express $\kappa$ and $\eta$ 
through the non-equal time equilibrium correlation functions and show that in 
general the mass is related to virtual (off-shell) excitations while the drag 
coefficient is determined by real (on-shell) processes. The 
expansion in derivatives of $\dot\lambda_\mu$ used to derive 
Eqs.~(\ref{emergent}) relies on the time scale separation between slow dynamics 
of the microscopic parameters $\vec\lambda$ and the fast dynamics of the system 
coupled to $\vec\lambda$. In the situations where this time scale separation 
does not hold, e.g. near phase transitions or low dimensional gapless system the 
validity of the expansion in time derivatives of $\vec\lambda$ should be checked 
on the case by case basis (see Sec. \ref{coriolis}).

Our result for the drag coefficient in Eq.~\eqref{emergent} agrees with the 
recent classical result of Ref.~\cite{sivak_2012} and, as shown there, can be 
used to define the dissipative metric. In addition we find that in the 
high-temperature or classical limit the mass tensor $\kappa$ is equal to the 
product of the inverse temperature and the thermally averaged Fubini-Study 
metric tensor $g_{\mu\nu}$. The latter characterizes deformations of the 
eigenstates of the systems to the change of 
$\vec\lambda$~\cite{provost_1980,venuti_2007,kolodrubetz_2013}:
\be
\kappa_{\nu\mu}=\beta g_{\nu\mu}
\label{eq_kappa}
\ee
This result allows one to interpret the (inertial) mass as an emergent property 
deriving from the response of the Hilbert space to the change of $\vec\lambda$. 
This statement echoes the result in general relativity in which the mass is seen 
as an emergent property which stems from the curvature of space-time. 
Let us point that the metric tensor $g_{\mu\nu}$ is equal to the covariance 
matrix of the gauge potentials $\mathcal A_\mu=i\partial_\mu$~\footnote{Except where explicitly mentioned  we work in the units $\hbar=1$.} responsible for 
the translations in the parameter space~\cite{kolodrubetz_2013}: 
$2g_{\mu\nu}=\langle \mathcal A_\mu \mathcal A_\nu\rangle_c+\mu\leftrightarrow 
\nu\equiv\langle \mathcal A_\nu \mathcal A_\mu\rangle-\langle \mathcal 
A_\nu\rangle \langle \mathcal A_\mu\rangle+\mu\leftrightarrow\nu$, where 
$\langle\dots\rangle$ implies averaging with respect to the equilibrium density 
matrix.
For translations in real space this gauge potential is nothing but the momentum 
operator $\mathcal A_x=p_x$, for rotations in $xy$ plain it is the angular 
momentum operator $\mathcal A_\phi=L_z$. In general these gauge potentials are 
complicated many-body operators, which however satisfy properties of 
locality~\cite{kolodrubetz_2013}. For simple setups like a free particle in a 
vacuum our result $m=\kappa_{xx}=\beta \langle p_x^2\rangle$ reduces to a 
conventional equipartition theorem $\langle p_x^2\rangle=m T$ (see Sec 
\ref{oscillator-example} for a detailed example). So this finding shows that 
even ordinary mass of a particle has a direct geometric interpretation and is 
related to the distortion of the Hilbert space surrounding the particle as it 
moves in space. At arbitrary temperatures instead of Eq.~(\ref{eq_kappa}) we 
obtain
\be
2\kappa_{\nu\mu}=\int_0^\beta d\tau\,\,\langle \mathcal A_{H,\nu}(-i\tau) 
\mathcal A_{H,\mu}(0)+\nu\leftrightarrow\mu\rangle_{0,c},
\label{eq_kappa1}
\ee
where $\mathcal A_{H,\mu}(-i\tau)$ is the imaginary time Heisenberg 
representation of the gauge potential. In the high-temperature limit 
Eq.~(\ref{eq_kappa1}) obviously reduces to Eq.~(\ref{eq_kappa}). This general 
expression can serve as the formal definition of the mass tensor for an 
arbitrary system. It also can be thought of as an extension of the equipartition 
theorem to quantum systems.

In less trivial situations of interacting systems and motion in the parameter 
space rather than the real space our results suggest an experimentally feasible 
way to indirectly measure the Fubini-Study metric and to directly analyze the 
quantum geometry of the system including various singularities and geometric 
invariants~\cite{kolodrubetz_2013}. We illustrate how this can be done by 
analyzing 
three simple examples (Sec.~\ref{examples}).

\section{Setup and main results}

We consider the following Hamiltonian: 
\begin{equation}
H_{tot}(\vec\lambda)=H_{0}(\vec{\lambda})+H(\vec{\lambda}),\label{eq:general}
\end{equation}
where $H_{0}(\vec{\lambda})$ is the Hamiltonian describing the bare motion
of the macroscopic d.o.f. $ \vec{\lambda}$, which can be
multi-component and $H(\vec{\lambda})$ is the Hamiltonian of the
interacting system of interest which depends on $\vec{\lambda}$. The choice of 
splitting $H_{tot}$ between $H_0$ and $H$ is somewhat arbitrary and we can well 
choose $H_0=0$ so that $H_{tot}=H$, however, for an intuitive interpretation of 
the results, it is convenient to assume that $H_{0}(\vec{\lambda})$ represents a 
massive degree of freedom in some external potential $V(\vec{\lambda})$: 
\[
H_{0}(\vec{\lambda})=\frac{\vec{p}_{\lambda}^{\,2}}{2m}+V(\vec{\lambda}).
\]
In the infinite mass limit ($m\rightarrow\infty$), $\vec{\lambda}$ 
represents an external (control) parameter whose dynamics is specified a priori. 
When $m$ is finite, $\vec{\lambda}$ is a dynamical variable and its dynamics 
needs to be determined self-consistently. 

One can identify two different sources of the energy change in the system. The 
first contribution is related to the dependence of the energy eigenstates of the 
system on $\vec{\lambda}$. This contribution does not vanish in the adiabatic 
limit and it is reversible. The second contribution is related to the 
excitations created in the system and, as we shall see, it contains both 
reversible and irreversible terms. To shorten the notation we term the first 
contribution as the adiabatic work and the second one as the heat.

Formally these two contributions can be defined as: 
\[
\partial_{t}\langle H\rangle(t)\equiv Tr\left\{ 
\partial_{t}\left(H(t)\rho(t)\right)\right\} =\dot{W}(t)+\dot{Q}(t)
\]
 where $\rho(t)$ is the density matrix and the adiabatic work rate, 
$\dot{W}(t)$, and the heat rate, $\dot{Q}(t)$,
are defined as: 
\begin{eqnarray}
&&\dot{W}(t)=\dot{\vec{\lambda}}\sum_{n}\frac{\partial\epsilon_{n}}{\partial\vec
{\lambda}}\,\rho_{n}(t)=\dot{\vec{\lambda}}\cdot\nabla_{\vec{\lambda}}E(\vec{
\lambda}) \equiv -\dot{\vec{\lambda}}\,\cdot M_{\vec{\lambda}}, 
\label{eq:work}\\
&&\dot{Q}(t)=\sum_{n}\dot{\rho}_{n}(t)\,\epsilon_{n},
\label{eq:heat}
\end{eqnarray}
where $E=\sum_{n}\epsilon_{n}\rho_{n}$ is the total energy of the system, 
$\rho_n(t)$ is the probability of occupying the $n$-th 
instantaneous energy eigenstate, 
$\epsilon_{n}$ is the instantaneous energy and 
$M_{\vec\lambda}=-\nabla_{\vec{\lambda}}E(\vec{\lambda})$ is the conventional 
generalized force which we introduced in Eq.~\eqref{emergent} \cite{mynote}.
This force appears, for example, in the Born-Oppenheimer approximation schemes 
in which the energy of the electrons, calculated for fixed ions' positions, acts 
as a potential for the (classical) motion of the ions.

In the strictly adiabatic limit, according to the quantum adiabatic 
theorem, there are no transitions between the instantaneous energy levels and 
$\dot{Q}=0$.~\footnote{For this paper we put aside the question of what happens 
when the system is macroscopic and ergodic so that the level spacings are 
exponentially small.} Using time-dependent perturbation theory 
to second order in $\dot{\lambda}$, we have computed $\dot{\rho}(t)$, and 
obtained the leading non-adiabatic contributions to the heat production rate 
(see Appendix~\ref{sec:Derivation}):
\be
\begin{split}
\dot{Q}(t)&=\dot{\lambda}_{\nu}(t)\int_{0}^{t}dt'\int_{0}^{\beta}d\tau\,\dot{
\lambda}_{\mu}(t-t')\langle \mathcal M_{H,\nu}(t)\mathcal 
M_{H,\mu}(t-t'+i\tau)\rangle_{0,c} \\
&=\dot{\lambda}_\nu(t) \int_0^t dt' \,\dot\lambda_{\mu}(t-t')   \sum_{n\neq m} 
\frac{\rho^0_n-\rho^0_m}{\epsilon_m-\epsilon_n} e^{i(\epsilon_m-\epsilon_n)t'} 
\langle m_\lambda | \mathcal M_{\nu} |n_\lambda \rangle \langle n_\lambda 
|\mathcal M_{\mu} |m_\lambda\rangle
\label{main}
\end{split}
\ee
where we imply summation over repeated indices,  $|n_\lambda\rangle$ denote
the (many-body) eigenstates of the Hamiltonian $H(\vec{\lambda}(t))$  with 
energies $\epsilon_n$, $\rho^0_n$ is the initial stationary occupation 
probability which, for the most of the paper, we assume to be thermal 
$\rho_n=\exp[-\beta \epsilon_n]/Z$, $c$ implies the connected part of the 
correlation function and
\begin{equation}
\mathcal M_{H,\nu}(t)=-\exp[iHt]\left(\partial_{\lambda_{\nu}}H\right)\exp[-iHt]
\label{heisenberg_rep}
\end{equation}
is the Heisenberg representation (for {\em time-independent} Hamiltonian) 
of the conjugate force coupled to $\lambda_{\nu}$. 
The equilibrium expectation value of this force $\langle \mathcal 
M_{\lambda}\rangle_0=M_{\lambda}=-\nabla_{\vec{\lambda}}E(\vec{\lambda})$ is the 
conventional generalized  force introduced earlier. Note that $\epsilon_{n}$ and 
$|n\rangle$ are the many-body levels 
and many-body eigenstates of the full interacting system (including the bath if 
it is present). 
If the system is weakly coupled to the bath then Eq.~\eqref{main} can be 
simplified using standard tools of the many-body perturbation 
theory~\cite{mahan}. For simplicity in Eq.~\eqref{main} 
and~\eqref{heisenberg_rep} we assume that the energy spectrum $\epsilon_n$ does 
not change with $\vec\lambda$ but in the Appendix~\ref{sec:Derivation} we show a 
complete derivation without such an assumption. Then essentially by energies 
$\epsilon_n$ and the matrix elements one needs to understand instantaneous 
values taken at time $t$ (see Appendix~\ref{sec:Derivation}).
Eq.~(\ref{main}) applies to arbitrary times $t$ and as such describes both 
transient and
the long time regimes. It is clear that if $t$ becomes longer
than the relaxation time then the heat production rate becomes insensitive
to the initial time $t=0$ but, for small isolated systems and at low
temperatures, transients can be important (see Sec. \ref{spin-example} for an 
example). This
expression generalizes an earlier result by D. A. Sivak and G. E.
Crooks~\cite{sivak_2012} to quantum systems at arbitrary temperatures.
In the high temperature (or classical) limit the integration
over the imaginary time component $\tau$ reduces to a factor of the
inverse temperature $\beta$ and we recover Eq.~(11) from Ref.~\cite{sivak_2012}.

If the rate $\dot{\vec{\lambda}}$ changes slowly in time on the scales
determined by the relaxation rate in the system then we can expand
Eq.~(\ref{main}) to the leading order in time derivatives of $\vec{\lambda}$. 
For systems with the Hamiltonian $H$ obeying time-reversal symmetry we get 
(terms proportional to higher time derivatives of $\vec\lambda$ are discussed in 
Appendix \ref{sec:time-derivatives}) 
\begin{equation}
\dot{Q}(t)=\dot{\lambda}_{\nu}\eta_{\nu\mu}\dot{\lambda}_{\mu}+\dot{\lambda}_{
\nu}\kappa_{\nu\mu}\ddot{\lambda}_{\mu}
\label{main1}
\end{equation}
The first term represents the usual drag or friction tensor and the
second term (as we will see below) amounts to the mass renormalization
of the external parameter. At zero temperature the first dissipative term 
generically vanishes as
required by the fluctuation-dissipation relations while the second
contribution remains finite. Both tensors $\eta$ and $\kappa$ are symmetric 
under $\nu\leftrightarrow\mu$ and positive semi-definite (assuming positive 
temperature). 
Before describing the friction and mass tensors in details we discuss some 
consequences of our main result \eqref{main1}.

\section{Implications of Eq.(\ref{main1}).}

\subsection{Energy absorption in a driven system \label{energy_absorption}}
As the first implication of our results, 
we discuss qualitative features of energy absorption in
systems driven by the external parameter $\vec\lambda(t)$. We
will be interested in protocols longer than the relaxation time in
the system such that susceptibilities $\eta$ and $\kappa$ become
effectively time independent (at shorter times one expects transients,
which can be analyzed using the short time expansion of Eq.~(\ref{main})).
To avoid possible singularities we assume that the protocol starts
smoothly in time, i.e. $\dot{\vec\lambda}(0)=0$. Integrating Eq.~(\ref{main1})
we find
\begin{equation}
Q(t)\approx\eta_{\mu\nu}\int_{0}^{t}dt'\dot{\lambda_\mu}(t')\dot 
\lambda_\nu(t')+\frac{\kappa_{\mu\nu}}{2}
\dot\lambda_\mu(t)\dot\lambda_\nu(t)+\dots
\label{sum_integrate}
\end{equation}
From this expression we see that indeed $\eta$ plays the role of
dissipation while $\kappa$ plays the role of the additional mass
associated with the dressing of the parameter $\lambda$ by excitations.
It is clear that the first (dissipative) term gives a contribution
proportional to the total duration of the process and it is dominant at long 
times $t\to\infty$.
However, at low temperatures or nearly isolated small systems the coefficient
$\eta$ can be very small and the second (mass) term can be dominant
for long times. In particular, if $\dot{\lambda}(t)$ approaches a constant 
value, after an initial transient, 
$Q(t)$ will display a plateau for $t\lesssim t^{\ast}=\kappa/\eta$.  

\subsection{Emergent equations of motion}

Next we consider $\vec \lambda$ to be a dynamic d.o.f. describing a 
``particle'' coupled to a many-body system
and study its dynamics. 

\subsubsection{Dynamics of a scalar particle.} 

First let us assume that $\vec\lambda$ is a single component scalar object. The 
Hamiltonian is given by Eq.~(\ref{eq:general}). Noting that the total energy of 
the particle and the system is conserved and using Eq.~\eqref{eq:work} and 
Eq.~\eqref{main1} we find
\be
\begin{split}
&\frac{d}{dt}\left[\frac{m\dot{\lambda}^{2}}{2}+V(\lambda)+E(\lambda)\right]=0\,
\Rightarrow\,m\dot{\lambda}\ddot{\lambda}+\dot{\lambda}\partial_{\lambda}V+\dot{
W}+\dot{Q}=0, \\
&\Rightarrow\,m\dot{\lambda}\ddot{\lambda}+\dot{\lambda}\partial_{\lambda}V-\dot
{\lambda}M_{\lambda}+\eta\dot{\lambda}^{2}+\kappa\dot{\lambda}\ddot{\lambda}=0.
\label{eq:cons}
\end{split}
\ee
Dividing this equation by $\dot{\lambda}$ we find 
\begin{equation}
(m+\kappa)\ddot{\lambda}+\dot{\lambda}\eta=-\frac{\partial 
V}{\partial\lambda}+M_{\lambda}
\label{eq:damped_eom}
\end{equation}
i.e. we find that near the slow limit the dynamics of $\lambda$ is given by the 
classical Newton's equations of motion with an additional 
dissipative term. 
Note that the Newtonian form of the equation of 
motion~\eqref{eq:damped_eom} was not postulated.  It rather emerges from the 
leading non-adiabatic expansion of the density matrix describing the system 
coupled to $\vec \lambda$. Thus the effect of the system on the dynamics of 
the macroscopic d.o.f. reduces to three  contributions: mass renormalization 
($\kappa$), drag coefficient ($\eta$), and an extra force ($M_{\lambda}$). 
As mentioned earlier, $M_{\lambda}$ is the force which appears in 
the Born-Oppenheimer approximation in which 
the only effect of the system is to lead to an effective (conservative) 
potential for the motion of the macroscopic d.o.f..
For example in this approximation schemes the electrons in a solid lead to an 
effective potential for the motion of the ions.
By considering the non-adiabatic correction to the energy of the system, which 
we named heat (see Eq.~\eqref{main}),
we go beyond the Born-Oppenheimer approximation and find the emergent Newton's 
equation~\eqref{eq:damped_eom}.
Note that in general $\kappa$, $\eta$ and $M_\lambda$ depend on $\lambda$ and 
Eq.~\eqref{eq:damped_eom} can predict non-trivial dynamics. As we pointed 
earlier the Hamiltonian $H_0$ can be absorbed into the definition of $H$. In 
this case $m=0$ and $\partial_\lambda V=0$ and Eq.~\eqref{eq:damped_eom} reduces 
to the scalar version of Eq.~\eqref{emergent}. Moreover when $H_0=0$ the total 
mass and the total force are entirely determined by the interactions with the 
system. Let us note that when $\kappa$ depends on $\vec\lambda$ it is more 
accurate to write the mass term in Eq.~\eqref{eq:damped_eom} as 
$\frac{d}{dt}(\kappa\dot \lambda)$. The difference between this term and $\kappa 
\ddot\lambda$ is of order $\dot\lambda^2$ and it is beyond the accuracy of our 
expansion. However a more careful analysis shows that this term is energy 
conserving, which implies that it comes from the full derivative of 
$\kappa\dot\lambda^2$.

\subsubsection{Berry curvature and Coriolis force.\label{coriolis}}  

The previous derivation of the equations of motion was based on using the 
conservation of the total energy conservation leading to Eq.~(\ref{eq:cons}) and 
dividing it by $\dot\lambda$. For a multicomponent parameter energy conservation 
is not sufficient to fix the equations of motion, since, as it is well known in 
the case of rotations or magnetic field, the Coriolis or the Lorentz forces 
affect the dynamics but not the energy. To proceed within the same framework of 
non-adiabatic response we can evaluate the expectation value of the generalized 
force $\langle \mathcal M_\nu\rangle(t)\equiv Tr\left[ \rho(t) \mathcal 
M_\nu\right]$. We give the details of the derivation in 
Appendix~\ref{sec:Derivation} and here only quote the final result obtained 
using the same approximation as Eq.~(\ref{main}):
\be
\langle \mathcal M_\nu\rangle(t) = M_\nu-\int_0^t 
dt'\dot\lambda_\mu(t-t')\,\sum_{n\ne m} 
\frac{\rho_n^0-\rho_m^0}{\epsilon_m-\epsilon_n} \mathrm e^{i (\epsilon_m - 
\epsilon_n) t'} \langle m_\lambda|\mathcal M_\nu|n_\lambda\rangle\langle 
n_\lambda|\mathcal M_\mu|m_\lambda\rangle.
\label{main_M}
\ee
We remind that the generalized force $M_\nu$ is by definition the equilibrium 
expectation value of $\mathcal M_\nu$. If we now do the expansion of $\dot 
\lambda_\mu(t-t')$ near $t'=0$ up to the second derivative we find \cite{sign}
\be
\langle \mathcal 
M_\nu(t)\rangle=M_\nu+F_{\nu\mu}\dot\lambda_\mu-\eta_{\nu\mu}\dot\lambda_\mu
-\kappa_{\nu\mu}\ddot \lambda_\mu-F'_{\nu\mu}\ddot \lambda_\mu,
\label{off_diag}
\ee
where 
\be
F_{\nu\mu}=i\langle [\mathcal A_\nu,\mathcal A_\mu]\rangle=i\sum_n \rho^0_n 
\langle 
n_\lambda|\overleftarrow{\partial}_\nu\overrightarrow{\partial}
_\mu-\overleftarrow{\partial}_\mu\overrightarrow{\partial}
_\nu|n_\lambda\rangle\label{berry_c}
\ee
is the Berry curvature and
\be
F'_{\mu\nu}=i \pi \sum_{n\neq 
m}\,\frac{\rho_n^0-\rho_m^0}{\epsilon_n-\epsilon_m} \langle 
m_\lambda|\mathcal{M}_{\nu}|n_\lambda\rangle \langle 
n_\lambda|\mathcal{M}_{\mu}|m_\lambda\rangle\,\delta'(\epsilon_n-\epsilon_m)
\ee
is another on-shell antisymmetric tensor ($\delta'(x)$ stands for the derivative 
of the delta-function) and $\eta$ and $\kappa$ are the friction and the mass 
tensors discussed before. Without the acceleration terms Eq.~(\ref{off_diag}) 
extends earlier results on the dynamical Hall 
response~\cite{Viola1,Viola2,gritsev_2012} to finite temperatures and possibly 
open systems without extra assumptions about the Lindbladian dynamics 
\cite{avron_2011}. As in the familiar situation of the current response to the 
electric field (time dependent vector potential) the transverse Hall like 
response determined by the Berry curvature is non-dissipative (off-shell) while 
the longitudinal on-shell response proportional to the drag $\eta$ is directly 
related to dissipation~\cite{mahan}.

We can now self-consistently combine Eq.~(\ref{off_diag}) with the classical 
(Lagrangian) equations of motion for the parameter $\vec\lambda$:
\be
m_{\nu\mu} \frac{d\lambda_\nu}{dt}=p_\nu,\quad \frac{dp_\nu}{dt}=-\frac{\partial 
V}{\partial\lambda_\nu}+\langle\mathcal M_\nu(t)\rangle
\label{Hamilton}
\ee
and get the multicomponent dissipative Newton's equations:
\be
(m_{\nu\mu}+\kappa_{\nu\mu}+F'_{\nu\mu})\ddot \lambda_\mu+(\eta_{\nu\mu} 
-F_{\nu\mu})\dot\lambda_\mu=-\frac{\partial V}{\partial\lambda_\nu}+M_\nu.
\label{Newton}
\ee
which generalize Eq.~\eqref{eq:damped_eom}. 
The first term in this equation represents the renormalized mass (as we 
discussed we can choose the bare mass to be zero by absorbing $H_0$ into $H$). 
The term $\eta_{\nu\mu}\dot\lambda_\mu$ is the dissipative term also appearing 
in the single component case. And finally there are two new antisymmetric terms, 
one proportional to the Berry curvature is clearly the analogous of the Coriolis 
force and the other antisymmetric on-shell contribution encoded in $F'$ is 
effectively antisymmetric mass term. 
This suggests that the Berry curvature can be indirectly measured via the 
Coriolis force acting upon a macroscopic d.o.f.~\cite{measure_berry}. In systems 
with time-reversal symmetry the two tensors $F,F'$ vanish \cite{FFprime} and 
therefore, by fitting the long time dynamics of $\vec{\lambda}(t)$ to 
Eq.~\eqref{Newton} with the additional knowledge of equilibrium generalized 
force one can extract both the drag tensor, $\eta_{\nu\mu}$, and the mass 
tensor, $\kappa_{\nu\mu}$. We emphasize again that, at high temperatures, the 
mass tensor reduces to the Fubini-Study metric tensor which is a covariance 
matrix of the gauge potential, or the momentum operator, which translates the 
Hamiltonian eigenstates in the parameter space. Therefore the very notion of the 
mass is related to the distance between eigenstates of the Hamiltonian 
$H(\vec\lambda)$ induced by the  change of the parameter $\vec\lambda$. 
Let us point that Eq.~(\ref{main_kappa}) [see below] essentially guarantees UV 
convergence of the mass as long as the variance of the generalized force is 
finite. On the contrary in gapless systems or near singularities like continuous 
phase transitions or glassy systems where correlation functions slowly decay in 
time the mass can acquire infra-red divergences and the energy absorption 
becomes non-analytic function of the rate as discussed in Ref.~\cite{pg_2008}. 

Equation~\eqref{Newton} has another interesting implication. At 
zero temperature both dissipative tensors ($\eta$ and $F'$) vanish (unless the 
system is tuned to a critical point or if it has gapless low-dimensional 
excitations~\cite{pg_2008}). In this case Eq.~\eqref{Newton} can be viewed as a 
Lagrangian equations of motion. 
If fact, it is easy to see that the Lagrangian: 
\be
\mathcal{L}=\frac{1}{2}\,\dot\lambda_\nu\,(m+\kappa)_{\nu\mu}\,\dot\lambda_\mu + 
\dot\lambda_\mu\,A_{\mu}(\vec\lambda)-V(\vec\lambda)-E_0(\vec\lambda)
\label{lagrangian}
\ee
reproduces Eq.~\eqref{Newton} where $A_{\mu}(\vec\lambda)=\langle 
0_\lambda|\mathcal A_\mu|0_\lambda\rangle$ and 
$E_0(\vec\lambda)=\langle 0_\lambda|H(\vec\lambda)|0_\lambda\rangle$ are the 
value of the Berry connection and Hamiltonian (see~\eqref{eq:general}) in the 
instantaneous ($\vec\lambda$-dependent) ground state and we have used (see 
Eq.~\eqref{berry_c}):
\[
\frac{\partial A_\mu}{\partial \lambda_\nu}-\frac{\partial A_\nu}{\partial 
\lambda_\mu}
=i\left[ \partial_\nu \langle 0_\lambda|\partial_\mu 
0_\lambda\rangle-\partial_\mu \langle 0_\lambda|\partial_\nu 0_\lambda\rangle 
\right]
=i \langle 
0_\lambda|\overleftarrow{\partial}_\nu\overrightarrow{\partial}
_\mu-\overleftarrow{\partial}_\mu\overrightarrow{\partial}_\nu|0_\lambda\rangle 
= F_{\nu\mu}.
\]
From the Lagrangian~\eqref{lagrangian} we can define the canonical momenta 
conjugate to the coordinates $\lambda_\nu$:
\be
p_\nu\equiv\frac{\partial\mathcal{L}}{\partial 
\dot{\lambda}_\nu}=(m_{\nu\mu}+\kappa_{\nu\mu})\dot{\lambda}
_\mu+A_\nu(\vec\lambda)
\ee
and the emergent Hamiltonian:
\be
\mathcal{H}\equiv\dot\lambda_\nu\,p_\nu-\mathcal{L}=\frac{1}{2} (p_\nu-A_\nu) 
(m+\kappa)_{\nu\mu}^{-1} (p_\mu-A_\mu)+V(\vec\lambda) +E_0(\vec\lambda). 
\ee
Clearly the Berry connection term plays the role of the vector potential. Thus we see that the whole formalism of the Hamiltonian dynamics for arbitrary macroscopic degrees of freedom is actually emergent.  Without mass renormalization this Hamiltonian was first derived in Ref.~\cite{berry_vector_potential} 
in which it was also shown that when the slow d.o.f. is quantum, there is an addittional force proportional to 
the Fubini-Study metric tensor $g_{\nu\mu}$. Away from the ground state the dissipative tensors ($\eta$ and $F'$) are, in  general, non-zero and it is not possible to  reformulate Eq.~\eqref{Newton} via Hamiltonian dynamics.

\subsubsection{Dynamics of a conserved degree of freedom. Emergent equilibrium 
from dynamics.}

It is straightforward to apply the results above to the setup where two systems 
are coupled by a single conserved degree of freedom, i.e. 
$H=H_1(\lambda_1)+H_2(\lambda_2)$ with the additional constraint 
$\lambda_1+\lambda_2={\rm const}$. Then using the additivity of the mass 
$\kappa$ and drag $\eta$, which are obvious from the microscopic expressions 
(see Eq.~\eqref{eq:main-eta} and \eqref{main_kappa} below) and invariance of 
Eq.~(\ref{main1}) under $\lambda\to -\lambda$, we obtain the dissipative 
non-Markovian dynamics for $\lambda_1(t)$:
\begin{eqnarray}
(\kappa_1+\kappa_2) \ddot\lambda_1+(\eta_1+\eta_2) \dot 
\lambda_1=M^{(1)}_{\lambda_1}-M^{(2)}_{\lambda_2}
\label{conserved_quantity}
\end{eqnarray}
We note that the Markovian limit of Eq.~\eqref{conserved_quantity} is obtained 
by setting $\kappa_1=\kappa_2=0$.
As expected from basic thermodynamics, the equilibrium between two systems 
($\lambda_1=\text{const}$) is only possible when the generalized forces between 
two systems are equal to each other. Unlike in statistical mechanics, where this 
statement follows from the maximum entropy principle, here we explicitly derive 
this condition from the dynamical equilibrium. Note that this condition for the 
dynamical equilibrium does not require the two systems to be ergodic. In our 
derivation we only relied on starting from a stationary state, thermal or not.

\section{Detailed description of friction and mass tensor}

We now discuss the friction and mass tensors in details. 
The friction tensor can be expressed as \cite{Hanggi}
\begin{equation}
\begin{split}
\eta_{\nu\mu}&=\frac{1}{2}\int_{0}^{t}\, dt'\int_{0}^{\beta}\, d\tau\langle 
\mathcal M_{H,\nu}(t'-i\tau)\mathcal 
M_{H,\mu}(0)+\nu\leftrightarrow\mu\rangle_{c}\\
&=\sum_{n\neq 
m}\,\,\frac{\rho^0_{n}-\rho^0_{m}}{\epsilon_{m}-\epsilon_{n}}\,\langle 
m_\lambda|\mathcal M_{\nu}|n_\lambda\rangle\langle n_\lambda|\mathcal 
M_{\mu}|m_\lambda\rangle\,\frac{\sin\left((\epsilon_{m}-\epsilon_{n})t\right)}{
\epsilon_{m}-\epsilon_{n}}
\label{eq:main-eta}
\end{split}
\end{equation}
from which it is clear that at positive temperatures (or more generally for any 
passive density matrix such that 
$(\rho^0_{n}-\rho^0_{m})(\epsilon_{m}-\epsilon_{n})>0$) the friction tensor 
is positive semi-definite and can therefore be used to define a metric. 
In particular, the metric associated with the friction $\eta$ was used in 
Ref.~\cite{zulkowski_2012}
for finding the paths of minimal dissipation in the parameter space.
At long times, $t\rightarrow\infty$, we have
\[
\frac{\sin(\epsilon_{m}-\epsilon_{n})t}{\epsilon_{m}-\epsilon_{n}}
\to\pi\delta(\epsilon_{n}-\epsilon_{m}),\;\frac{\rho^0_{n}-\rho^0_{m}}{\epsilon_
{m}-\epsilon_{n}}\,\to\beta\,\rho^0_{m}.
\]
Therefore the friction tensor is defined by the on-shell processes, which is 
expected since we are dealing with slow, zero frequency, driving
\[
\eta_{\nu\mu}=\pi\beta\sum_{n\neq m}\,\,\rho^0_{m}\langle m_\lambda|\mathcal 
M_{\nu}|n_\lambda\rangle\langle n_\lambda|\mathcal 
M_{\mu}|m_\lambda\rangle\delta(\epsilon_{n}-\epsilon_{m}).
\]
So at $t\to\infty$ the drag is always determined by the high temperature 
asymptote of Eq.~(\ref{main}) (and thus the result of Ref.~\cite{sivak_2012} 
always holds). At finite times, the expression for $\eta$ has different high 
temperature and low temperature asymptotes. 

Similarly we can analyze the mass tensor $\kappa$  
\begin{equation}
\kappa_{\nu\mu}=-\frac{1}{2}\int_{0}^{t}\, dt'\, t'\int_{0}^{\beta}\, 
d\tau\langle \mathcal M_{H,\nu}(t'-i\tau)\mathcal 
M_{H,\mu}(0)+\nu\leftrightarrow\mu\rangle_{c},
\label{kappa_t}
\end{equation}
which in the long time limit can be written as
\be
\begin{split}
\kappa_{\nu\mu}&=\sum_{n\neq 
m}\frac{\rho^0_{n}-\rho^0_{m}}{\left(\epsilon_{m}-\epsilon_{n}\right)^{3}}\,
\langle m_\lambda|\mathcal M_{\nu}|n_\lambda\rangle\langle n_\lambda|\mathcal 
M_{\mu}|m_\lambda\rangle \\
&=\sum_{n\neq m}\frac{\rho^0_{n}-\rho^0_{m}}{\epsilon_{m}-\epsilon_{n}}\,\langle 
m_\lambda|\overleftarrow{\partial_{\lambda_{\nu}}}|n_\lambda\rangle\langle 
n_\lambda|\overrightarrow{\partial_{\lambda_{\mu}}}|m_\lambda\rangle
\label{main_kappa}
\end{split}
\ee
where we have used the identity (see e.g. Ref.~\cite{degrandi_2013}) 
\begin{equation*}
\left( \epsilon_{m}-\epsilon_{n}\right)\langle 
n_\lambda|\overrightarrow{\partial_{\lambda_{\mu}}}|m_\lambda\rangle=\langle 
n_\lambda|\mathcal M_{\mu}|m_\lambda\rangle
\end{equation*}
We emphasize again that we put aside the issue of convergence of 
the sum in Eq.~(\ref{main_kappa}) when we deal with macroscopic ergodic systems 
at finite temperatures. This convergence is essentially guaranteed if the 
non-equal time correlation functions entering Eq.~(\ref{kappa_t}) decay fast in 
time, which was our key assumption. Unlike for the friction tensor $\eta$, the 
mass tensor $\kappa_{\nu\mu}$  has, in the infinite time limit, non-vanishing 
asymptotes both in the high and zero-temperature limits. At low temperatures, 
$\beta\to\infty$,
\be
\kappa_{\nu\mu}\approx\sum_{m\ne0}\frac{\langle0_\lambda|\mathcal 
M_{\nu}|m_\lambda\rangle\langle m_\lambda|\mathcal 
M_{\mu}|0_\lambda\rangle+\nu\leftrightarrow\mu}{\left(\epsilon_{m}-\epsilon_{0}
\right)^{3}}\label{kappa_low}.
\ee
At high temperatures (or near the classical limit) we find
\be
\kappa_{\nu\mu}\approx\frac{\beta}{2}\sum_{m}\rho_{m}\,\left(\langle 
m_\lambda|\overleftarrow{\partial_{\lambda_{\nu}}}\,\overrightarrow{\partial_{
\lambda_{\mu}}}|m_\lambda\rangle_c+\nu\leftrightarrow\mu\right)=\beta\, 
g_{\nu\mu}
\label{kappa_high}
\ee
where $g_{\nu\mu}$ is the finite temperature version Fubini-Study
metric tensor characterizing statistical average of the distance between
many-body eigenstates~\cite{kolodrubetz_2013}.  Let us mention that in traditional units the expressions for the mass Eqs.~(\ref{main_kappa}) - (\ref{kappa_high}) should be multiplied by $\hbar^2$, which follows from the correct definition of the Gauge potentials $i\partial_{\lambda_\mu}\to i\hbar \partial_{\lambda_{\mu}}$.  We also point that the mass tensor $\kappa$ can be 
written through the integrated connected imaginary time correlation function of 
the gauge potentials $\mathcal A_\nu$ and $\mathcal A_\mu$:
\be
\kappa_{\nu\mu}=\frac{1}{2}\int_0^\beta d\tau \langle \mathcal A_{H,\nu}(-i\tau) 
\mathcal A_{H,\mu}(0)+\nu\leftrightarrow\mu\rangle_{0,c}
\ee
At high temperatures the imaginary time integral reduces to a factor $\beta$ and 
this result clearly reduces to Eq.~(\ref{kappa_high}). We point that the 
expressions for $\eta$ in the second line of (\ref{eq:main-eta}) and for 
$\kappa$ in (\ref{main_kappa}) apply to arbitrary stationary distributions 
$\{\rho^0_{n}\}$, not necessarily thermal.

It is interesting to note that short time transient dynamics also
bears the geometric information. Thus integrating Eq.~(\ref{main})
over short times (and again assuming high temperature or classical
limit) gives 
\[
Q(t)\approx\frac{\beta}{2}\,\delta\lambda_{\nu}(t)\,{f}_{\nu\mu}\,\delta\lambda_
{\mu}(t),
\]
where $\delta\lambda_{\nu}(t)\equiv\lambda_{\nu}(t)-\lambda_{\nu}(0)$
and 
\begin{equation}
f_{\nu\mu}=\frac{1}{2}\langle \mathcal M_{\nu}\mathcal M_{\mu}+\mathcal 
M_{\mu}\mathcal 
M_{\nu}\rangle_{0,c}=\frac{\partial^{2}\ln(Z)}{\partial\lambda_{\mu}
\partial\lambda_{\nu}}
\label{short}
\end{equation}
is the thermodynamic metric tensor defined through the second derivative
matrix of the partition function~\cite{ruppeiner_1979,crooks_2007}.

\section{Examples\label{examples}}

\subsection{Mass Renormalization of a particle interacting with identical 
quantum oscillators.\label{oscillator-example}}

We now show explicitly how the mass of a classical object coupled to a quantum 
environment is renormalized. 
Let us consider the Hamiltonian \eqref{eq:general} where
\begin{equation*}
H_0=\frac{P_{\lambda}^2}{2\mu}+V(\lambda),\,\,
\,H(\lambda)=\sum_{j=1}^{N}\left( 
\frac{\hat{p}_j^2}{2m}+\frac{k}{2}(\hat{x}_j-\lambda)^2\right)+\sum_{i>j}U(\hat{
x}_i-\hat{x}_j)
\end{equation*} 
describes a macroscopic d.o.f. interacting with a collection of coupled quantum 
oscillators (QOs). 
We note that the coupling between the oscillators, $U(\hat{x}_i-\hat{x}_j)$, 
does not need to be harmonic so instead of oscillators we can deal with 
interacting particles. Here, for simplicity we consider the $1$-dimensional case 
and $\hat{x}_j$ and $\hat{p}_j$ are quantum operators satisfying 
the usual commutation relations $\left[\hat{x}_j,\hat{p}_l\right]=i 
\delta_{j,l}$ and we assume that the macroscopic d.o.f is 
subject to a constant external force, i.e. $-\partial_{\lambda}V(\lambda)=F_0$.

The basic expectation is that, due to the interaction between the macroscopic 
d.o.f. and the QOs, the macroscopic d.o.f. will drag the QOs with itself as it 
moves and as a consequence its mass will be renormalized. This expectation can 
be confirmed by analyzing the evolution of the center of mass of the system 
which, by definition, is determined only by the external forces $m_{eff} 
\ddot{X}_{cm}(t)=F_0$, where $m_{eff}=\mu+N m$ is the total mass. The position 
of the macroscopic d.o.f. is the vector sum of the position of the center of 
mass and the relative coordinate (which will in general have an oscillatory 
behavior) $\lambda(t)=X_{cm}(t)+X_{\lambda}(t)$. We therefore conclude that, up 
to the oscillations of the relative coordinate, the macroscopic d.o.f. evolves 
with the renormalized mass $m_{eff}=\mu+N m$ which can be several times larger 
than its bare mass $\mu$. If the coupling between oscillators is zero, $U=0$, 
the dynamics of the macroscopic d.o.f. can be solved exactly and it can be 
verified that our argument is correct and that the oscillations of the relative 
coordinate have frequency $\Omega\sim N^{1/2}$ and amplitude $B\sim N^{-1/2}$, 
which can be neglected for $N\gg1$.

We now show how this behavior is reproduced using the emergent Newton's law 
Eq.~\eqref{eq:damped_eom}. Let's first analyze the case of 
$N$ independent Harmonic Oscillators (HO), i.e. $U=0$. Then the drag term $\eta$ 
is zero since the harmonic oscillators have a discrete spectrum and there are 
not gapless excitations \cite{HObath}. Moreover the extra force term 
$M_{\lambda}\equiv\sum_n \rho_n \partial_{\lambda}\epsilon_n$ is also zero since 
$\epsilon_n=\omega (n+\frac{1}{2})$ is independent on $\lambda$. 
Therefore we only need the mass tensor $\kappa$ which can be computed via 
Eq.~\eqref{main_kappa} 
where $\mathcal{M}_{\lambda}\equiv -\partial_{\lambda}H=k(\hat{x}-\lambda)$. 
Expressing the position via the ladder operators 
$\hat{x}=\sqrt{\frac{1}{2m\omega}}(a+a^{\dagger})$ and using the standard 
properties $a|n\rangle=\sqrt{n}|n-1\rangle$ and 
$a^\dagger|n\rangle=\sqrt{n+1}|n+1\rangle$  together with the thermal 
occupations $\rho_n=e^{-\beta\omega n}\left(1-e^{-\beta\omega}\right)$ it is 
straightforward to show that $\kappa=m$. Since there are $N$ independent HOs 
each contributing equally to the renormalization of the mass 
Eq.~\eqref{eq:damped_eom} becomes:
\be
(\mu+N m)\ddot{\lambda}=F_0 
\ee
in perfect agreement with our discussion above. 

We can extend the above analysis to arbitrary interaction by using the 
equipartition theorem  (see Eq.~\eqref{main_kappa}): $\kappa=\beta \langle 
(\sum_j p_j) (\sum_l p_l) \rangle_0= \beta \langle p_{tot}^2 \rangle_0=N m$ 
where $p_{tot}=\sum_j p_j$ is the total momentum of the oscillators. Therefore 
even when $U\ne 0$ we find that the renormalized mass is $m_{eff}=\mu+N m$.

\subsection{Mass renormalization in a quantum piston.}

\begin{figure}
\includegraphics[width=0.90\columnwidth]{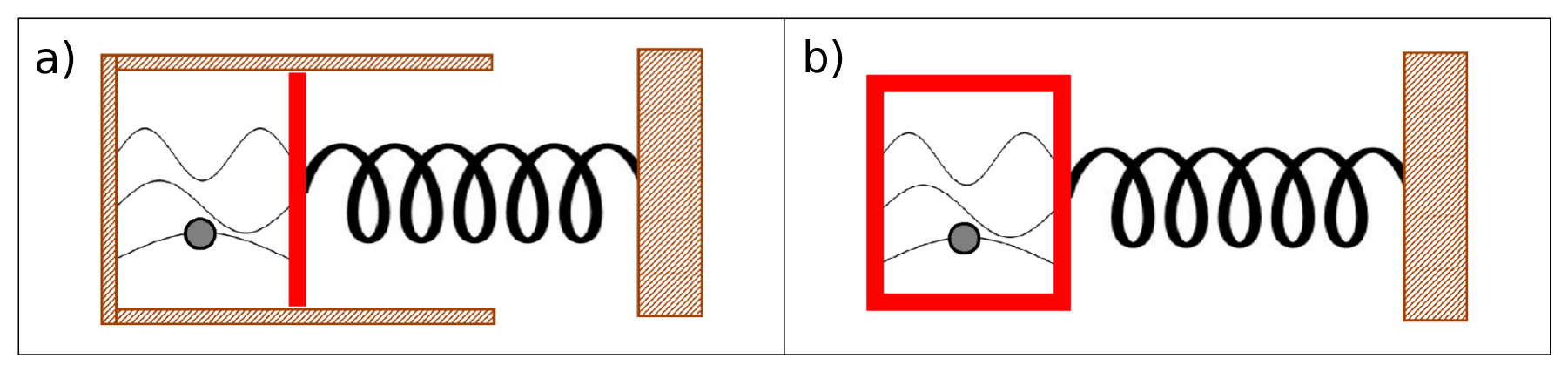}
\caption{(Color on-line) Schematic of a quantum piston. 
a) The spring is connected to a wall of the potential in which a quantum 
particle of mass $m$ is initially confined into the ground state.
b) As in a) but now the spring is connected to the whole potential well which 
moves rigidly.
The wave-lines in a) and b) represent the low energy wavefunctions of the 
quantum particle in the confining potential.}
\label{fig:piston}
\end{figure}

Let us consider a massless spring connected to the potential wall. In this section we will explicitly insert all factors of $\hbar$.
We also imagine that a quantum particle of mass $m$ is initially prepared in the ground state of the  confining potential
(see panel ``a" in Fig.~\ref{fig:piston}).
As in the previous example we will compute how the mass of a classical object 
(the spring) coupled to a quantum environment (the potential well) is 
renormalized.

According to Eq.~\eqref{kappa_low} the mass renormalization is given by
\be
\kappa_R=2 \hbar^2 \sum_{n\neq 0} \frac{|\langle n_\lambda| 
\mathcal{M}|0_\lambda\rangle|^2}{(\epsilon_n-\epsilon_0)^3},
\label{eq:kappa_R}
\ee
where $\lambda=X_R$ is the position of the right potential wall.
We approximate the confining potentail as a very deep square well potential. 
Then $\mathcal{M}\equiv-\partial_\lambda H=-V\delta(x-X_R)$ and we find
\be
\kappa_R=2 \hbar^2 \sum_{n\neq 0} \frac{V^2 |\psi_0(X_R)|^2 |\psi_n(X_R)|^2}{(\epsilon_n-\epsilon_0)^3}.
\ee
Using the well known result for a finite (and deep) square well potentail
\[
|\psi_n(X_R)|=\sqrt{\frac{2}{L}}\,\,\sqrt{\frac{\epsilon_n}{V}}
\]
where the factor of $\sqrt{2/L}$ comes from the normalizazion of the wave-function in a square potential of lenght $L$, we obtain
\be
\kappa_R=2 \hbar^2 \left(\frac{2}{L}\right)^2 \sum_{n\neq 0} \frac{\epsilon_0 \epsilon_n}{(\epsilon_n-\epsilon_0)^3}.
\label{KR}
\ee
Substituting 
\[
\epsilon_n = \frac{\hbar^2 k_n^2}{2m},\quad k_n=\frac{n+1}{L}\,\pi,\quad\forall n\ge0
\]
we arrive at
\[
\kappa_R=m\,\frac{16}{\pi^2}\sum_{n\geq 1} \frac{(n+1)^2}{[(n+1)^2-1]^3} 
=m\,\frac{2\pi^2-3}{6\pi^2}\approx 0.28 m
\]
The result is identical if we connect the piston to the left wall, i.e. $\kappa_L=\kappa_R$. 

Now let us consider a slightly different setup where the spring connects to the 
whole potential well 
(see panel ``b" in Fig.~\ref{fig:piston}) so that $\lambda$ now indicates the 
center of mass of the well.
From the Galilean invariance we expect $\kappa=m$. In fact, since now both 
potentials walls are moving, our expression gives
\[
\mathcal{M}=-\partial_\lambda H=-V (\delta(x-X_R)-\delta(x-X_L)),
\]
where $X_L$ and $X_R$ are the left and right positions of the walls. Thus using 
Eq.~(\ref{eq:kappa_R}) we obtain
\be
\kappa_+=2 \hbar^2 \sum_{n\neq 0} \frac{V^2(\psi_0(X_L)\psi_n(X_L)-\psi_0(X_R) \psi_n(X_R))^2}{(\epsilon_n-\epsilon_0)^3}
\label{eq:kappa+}
\ee
Since in a symmetric potential well $\psi_n(X_R)=(-1)^n\psi_n(X_L)$ only the odd terms contribute in the equation above.
Following the same line of reasoning as before we arrive at (note the extra factor of $4$ with respect to Eq.~\eqref{KR})
\[
\kappa_+=2 \hbar^2 \left(\frac{2}{L}\right)^2 4 \sum_{n=odd} \frac{\epsilon_0 \epsilon_n}{(\epsilon_n-\epsilon_0)^3}  \\
=m \frac{64}{\pi^2} \sum_{n=odd} \frac{(n+1)^2}{[(n+1)^2-1]^3}=m
\]
So indeed we recover the expected result. This simple calculation illustrates 
that indeed we can understand the notion of the mass as a result of virtual 
excitations created due to the acceleration of the external coupling (position 
of the wall(s) in this case). 
If instead we analyze the setup where the two walls are connected to a spring 
and move towards each other so that 
$\lambda$ is the (instantaneous) length of the potential well we find
\begin{equation}
\begin{split}
\kappa_-&=2 \hbar^2 \sum_{n\neq 0} \frac{V^2(\psi_0(X_L)\psi_n(X_L)+\psi_0(X_R) \psi_n(X_R))^2}{(\epsilon_n-\epsilon_0)^3}\\
&=m \frac{64}{\pi^2} \sum_{\substack{n=even \\ n\neq 0}} \frac{(n+1)^2}{[(n+1)^2-1]^3} = m \frac{\pi^2-6}{3\pi^2}\approx 0.13 m.
\label{eq:kappa-}
\end{split}
\end{equation}

Let us point another peculiar property of the mass. Clearly 
$\kappa_L+\kappa_R\approx 0.56 m\neq \kappa_+,\kappa_-$, i.e. the mass 
renormalization of the two walls is not the same as the sum of the mass 
renormalization of each wall measured separately. This is the result of quantum 
interference, which is apparent in Eqs.~\eqref{eq:kappa+} and \eqref{eq:kappa-}. 
Note that $(\kappa_+ + \kappa_-)/2=\kappa_L+\kappa_R$. Thus the mass behaves 
similarly to the intensity in the double pass interferometer, where the sum of 
intensities in the symmetric and antisymmetric channels is conserved. At a high 
temperature or in the classical limit the interference term will disappear and 
we will find $\kappa_L=\kappa_R\approx 0.5 m$.

\subsection{Energy absorption of a spin in a rotating magnetic 
field.\label{spin-example}}

\begin{figure}
\includegraphics[width=0.90\columnwidth]{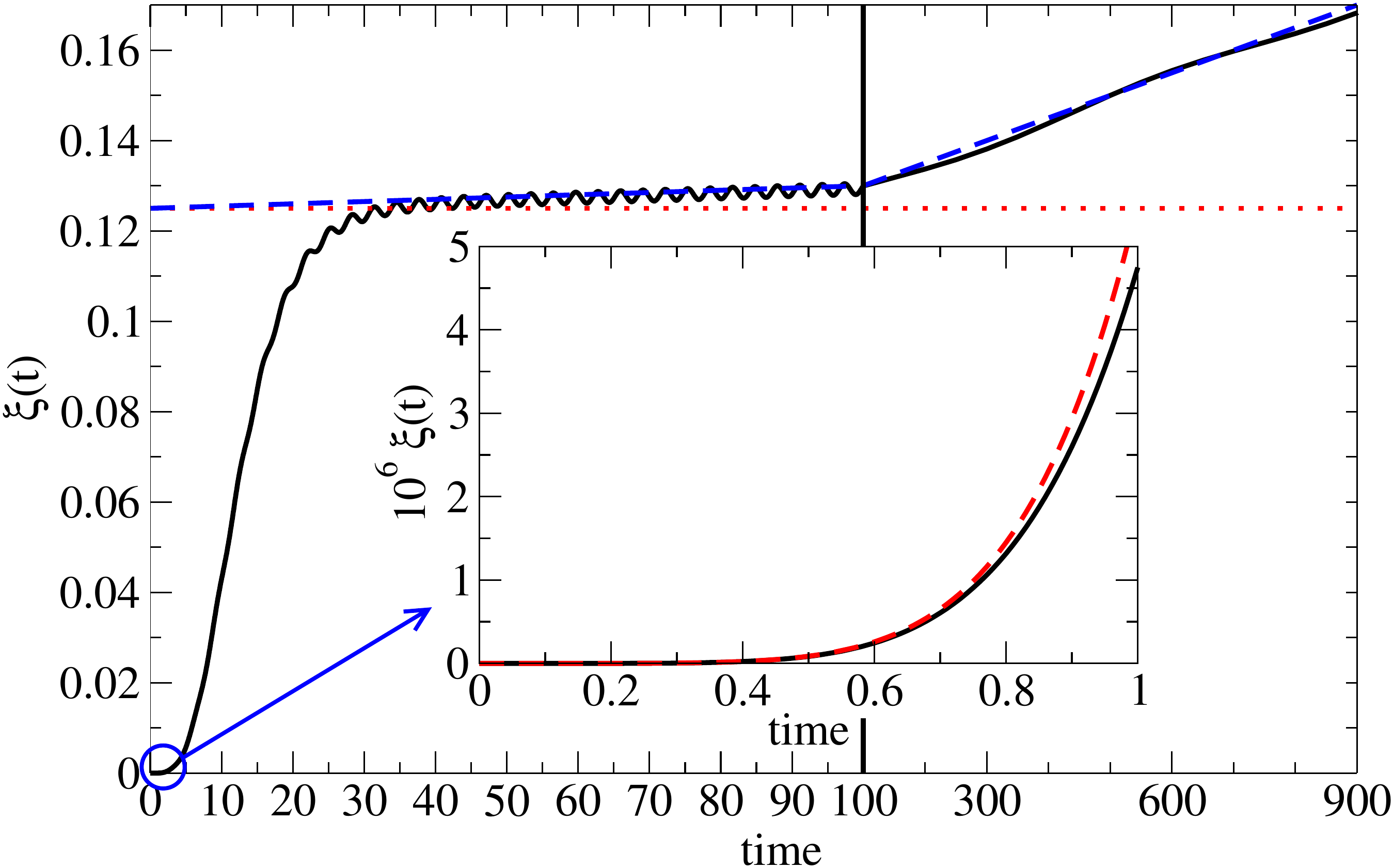}
\caption{(Color on-line) Rescaled heat, 
$\xi(t)\equiv\frac{Q(t)}{v_0^2\Delta\tanh(\beta\Delta)}$, 
for the protocol $\dot{\lambda}(t)$ discussed in the main text. 
(Main graph) At long time, $t\gg\tau_v,\Delta^{-1}$,the heat is well 
approximated by the sum of the mass renormalization 
and dissipative contributions (dashed blue line) (see Eq. 
\eqref{sum_integrate}). 
The horizontal (dotted red) line is the (constant) contribution from the mass 
renormalization.
(Inset) At short time, $t<\Delta^{-1}$ the heat is well approximated by 
$Q(t)=\Delta\tanh(\beta\Delta)\frac{(\delta\lambda(t))^2}{2}$ (see 
Eq.~\ref{short}) (red dashed line).
In both panels $\Delta=1,\,\tau_v=10,\,\tau_c=10^3$ so that the condition 
$\tau_c\gg\tau_v>\Delta^{-1}$ is well satisfied and the 
the plateau is well visible.}
\label{fig:spins}
\end{figure}

As a next example we consider a system of independent spin-$\frac{1}{2}$ in a
rotating magnetic field in the xz plane. Here we assume that the magnetic filed 
is an
external parameter whose dynamics is given a priori. First we assume that these
spins are isolated and later add a weak coupling to a bath. The Hamiltonian
of each of the spins reads 
\begin{equation}
H\left(\lambda(t)\right)=-\Delta\left(\cos(\lambda)\,\sigma_{z}+\sin(\lambda)\,
\sigma_{x}\right)+H_{SB}\label{eq:HH},
\end{equation}
where $H_{SB}$ is the part of the Hamiltonian representing possible coupling to 
the bath and which does not depend on the angle $\lambda$. The instantaneous 
eigenstates change during the dynamics described by the protocol $\lambda(t)$ 
while the energy levels remain unchanged.
The eigenstates of this Hamiltonian (excluding the bath) are trivially 
\begin{equation}
|gs_\lambda\rangle=\left(\begin{array}{c}
\cos\left(\frac{\lambda}{2}\right)\\
\sin\left(\frac{\lambda}{2}\right)
\end{array}\right),\quad|ex_\lambda\rangle=\left(\begin{array}{c}
\sin\left(\frac{\lambda}{2}\right)\\
-\cos\left(\frac{\lambda}{2}\right)
\end{array}\right)
\end{equation}
corresponding to the energies $E_{gs}=-\Delta$ and $E_{ex}=\Delta$
respectively.  It is straightforward to compute the mass $\kappa$ using 
Eq.~(\ref{main_kappa}): 
\begin{equation}
\kappa=\frac{\tanh(\beta\Delta)}{4\Delta} \
\label{kappa-simple}
\end{equation}
In the high temperature limit $\kappa\approx\beta/4$,
which is indeed a product of the inverse temperature and the single
component metric tensor, also known as the fidelity susceptibility
of the spin. Because this expression is non-singular, a small coupling to the 
bath can at most introduce small corrections to $\kappa$.  However, this is not 
the case with dissipative coefficient $\eta$, which
clearly vanishes in the long time limit without the bath because there are no 
gapless excitations. For typical coupling to the bath the transverse spin-spin 
correlation functions entering Eq.~(\ref{eq:main-eta}) oscillate with
frequency $2\Delta$ and decay with the relaxation time $\tau_c$ set by the 
bath~\cite{Lee,external_noise} so 
\be
\begin{split}
\eta&=\Delta\tanh(\beta\Delta)\int_{0}^{\infty}dt^{\prime}\cos(2\Delta 
t^{\prime})\exp\left[-\frac{t^{\prime}}{\tau_c}\right]\\
&=\tanh(\beta\Delta)\,\frac{\Delta\tau_c}{1+(2\Delta\tau_c)^{2}}\label{eq:eta-spins}
\end{split}
\ee
Within the same approximation the mass $\kappa$ is also modified:
\be
\begin{split}
\kappa&=\Delta\tanh(\beta\Delta)\int_{0}^{\infty}dt^{\prime}(-t^{\prime}
)\cos(2\Delta t^{\prime})\exp\left[-\frac{t^{\prime}}{\tau_c}\right]\\
&=\tanh(\beta\Delta)\,\Delta\tau_c^2\frac{(2\Delta\tau_c)^2-1}{
((2\Delta\tau_c)^2+1)^2}\label{eq:kappa-spins}
\end{split}
\ee
which reduces to Eq.~\eqref{kappa-simple} for $\tau_c\Delta\gg 1$. In the high 
temperature limit $\beta\Delta\ll 1$ this expression for the mass is equal to 
the inverse temperature times the metric tensor but now of the full system 
including bath degrees of freedom. 
These expressions Eq.\eqref{eq:eta-spins} and \eqref{eq:kappa-spins} do not 
apply for $\tau_c<\Delta^{-1}$ since the simple approximation to the 
exponentially decaying correlation function breaks in this limit due to the Zeno 
effect. Therefore we only consider the situation $\Delta\tau_c\gg1$. 
In Fig. \ref{fig:spins} we show the spin energy obtained by numerically 
integrating the expression
$\dot{Q}(t)=\dot{\lambda}(t)\int_0^t dt' \dot{\lambda}(t-t')\cos(2\Delta 
t')\exp[-t'/\tau_c]$ for the protocol $\dot{\lambda}(t)=v_0 \tanh^2(t/\tau_v)$. 
Note that $\dot{\lambda}(t)$ starts smoothly and approaches the constant value 
$v_0$ for $t\gtrsim 2\tau_v$. 
For $t<\Delta^{-1}$ the heat $Q(t)$ is well approximated by the short time 
expansion (see Eq.~\eqref{short}) while for $\Delta^{-1}<t\lesssim 
t^{\ast}=\kappa/\eta\approx \tau_c$ we observe a plateau in agreement with the 
general discussion in \ref{energy_absorption}.

\subsection{Central spin model. Extracting the Chern number from the Coriolis 
force.}

As a third example let us consider a macroscopic rotator (angular momentum) 
interacting with independent spin-$\frac{1}{2}$ particles. If instead of the 
rotator we use the quantum spin operator, this model is known as the central 
spin model. Because dynamics of a spin in any external field is always classical 
(the evolution of the Wigner function is exactly described by classical 
trajectories) there is no difference between quantum and classical dynamics in 
this case and we do not need to assume that the rotator is macroscopic. The 
Hamiltonian describing the system is \eqref{eq:general} with
\be
H_0=\frac{\vec L^2}{2I}+V(\vec{n}),\quad H=-\vec n \cdot \sum_{i=1}^{N} 
\Delta_i\,\vec \sigma_i,
\ee
where $I$ is the momentum of inertia, $V(\vec{n})$ is a time-dependent external 
potential, $\vec L$ is the angular momentum and $\vec n$ is the 
three-dimensional unit vector which can be parameterized using spherical angles: 
$\vec n=(n_x, n_y, n_z)=( \sin\theta\cos\phi,\sin\theta\sin\phi,\cos\theta)$. 
This example is similar to the previous one except that the effective magnetic 
field is no longer confined to the $xz$ plane and we no longer assume that it is 
given by an external protocol.
The time evolution of this system needs to be found self-consistently. On the 
one hand, each spin evolves according to the von Neumann equation with the 
time-dependent Hamiltonian $H(\vec{n}(t))$:
\begin{equation}
i\partial_t \rho=\left[H(\vec{n}(t)),\rho\right]
\label{vonN}
\end{equation} 
and, on the other hand, the rotator evolves according to the Hamilton equations 
of motion
\begin{equation}
I \dot{\vec{n}}=\vec L \times \vec n,\,\quad \dot{\vec{L}}=\vec n \times \left( 
\vec{M}_{ext} + \langle -\frac{\partial H}{\partial \vec n} \rangle \right)=\vec 
n \times \left( \vec{M}_{ext}+\sum_i \Delta_i \langle \vec \sigma_i 
\rangle\right)
\label{classical_rotor}
\end{equation} 
where $\vec{M}_{ext}=-\frac{\partial V(\vec n)}{\partial \vec n}$ is the 
external force and $\langle \dots \rangle$ indicates the quantum average over 
the density matrix $\rho(t)$ (see Eq.~\eqref{vonN}).
We assume that initially $\vec n_0=(0,0,1)$ and the spins are in thermal 
equilibrium with 
respect the Hamiltonian $H(\vec n_0)$, i.e. $\langle \sigma^x_i 
\rangle_0=\langle \sigma^y_i \rangle_0=0$ 
and $\langle \sigma^z_i\rangle_0= \tanh(\beta \Delta_i)$.

For the toy model proposed here these coupled equations can be easily solved 
numerically. In fact, according to the Ehrenfest theorem, the evolution of the 
expectation values follow the classical equation of motion and the von Neumann 
equation~\eqref{vonN} can be replaced with the much simpler equation $\dot{\vec 
m}_i=2\Delta_i\,\vec m_i \times \vec n$ where $\vec{m_i}=\langle \vec \sigma_i 
\rangle$. Therefore the exact dynamics of the system consists of the vectors 
($\vec L, \vec n, \{\vec m_i\})$ precessing around each other. 

We now compare the exact dynamics with the emergent Newton's law.
First, we note that the form of the equations~\eqref{classical_rotor} 
immediately implies
\begin{equation}
\begin{split}
&\dot{\vec n} \cdot \vec L =0, \,\, \vec n \cdot \dot{ \vec L}=0 \rightarrow 
\vec n \cdot \vec L=\text{const} \\
&\dot{\vec{n}}\cdot \vec n=0 \rightarrow |\vec 
n|^2=\text{const},\,\,\ddot{\vec{n}}\cdot \vec n=-|\dot{\vec n}|^2
\end{split}
\label{conservation}
\end{equation}
Next we need to compute the generalized force $\langle \vec{\mathcal{M}} 
\rangle=\langle -\frac{\partial H}{\partial \vec n} \rangle$, and the tensors 
$\kappa$ and $F$. The drag term $\eta$ and the anti-symmetric mass $F^\prime$ 
are zero since there are no gapless excitations. Therefore Eq.~\eqref{off_diag} 
reduces to:
\begin{equation*}
\langle \vec{\mathcal{M}} \rangle = \langle \vec{\mathcal{M}} 
\rangle_0+F_{\nu,\mu}\dot{n}_{\mu}-\kappa_{\nu,\mu}\ddot{n}_{\mu}
\end{equation*}
where $\nu,\mu=x,y,z$. The ground and excited states of each spin-$\frac{1}{2}$ 
are:
\be
|gs^i_{\theta,\phi}\rangle=\left(\begin{array}{c}
\cos\left(\frac{\theta}{2}\right)\mathrm e^{-i\phi/2}\\
\sin\left(\frac{\theta}{2}\right)\mathrm e^{+i\phi/2}
\end{array}\right),\quad
|ex^i_{\theta,\phi}\rangle=\left(\begin{array}{c}
\sin\left(\frac{\theta}{2}\right)\mathrm e^{-i\phi/2}\\
-\cos\left(\frac{\theta}{2}\right)\mathrm e^{+i\phi/2}
\end{array}\right)
\ee
with energy $\pm\Delta_i$ respectively from which it follows 
\begin{align*}
\langle \vec{\mathcal{M}} \rangle_0=0,\quad
&F=F_0 \left(\begin{array}{ccc}
0 & n_{z} & -n_{y}\\
-n_{z} & 0 & n_{x}\\
n_{y} & -n_{x} & 0
\end{array}\right),\\
&\kappa=\kappa_0 \left(\begin{array}{ccc}
1-n_{x}^{2} & -n_{x}n_{y} & -n_{x}n_{z}\\
-n_{y}n_{x} & 1-n_{y}^{2} & -n_{y}n_{z}\\
-n_{z}n_{x} & -n_{z}n_{y} & 1-n_{z}^{2}
\end{array}\right).
\end{align*}
where $F_0\equiv\frac{1}{2}\sum_i \tanh(\beta\Delta_i)$ and 
$\kappa_0\equiv\sum_i \frac{\tanh(\beta\Delta_i)}{4\Delta_i}$.
Substituting these expressions in Eq.~\eqref{classical_rotor} we find
\begin{equation*}
I \dot{\vec{n}}=\vec L \times \vec n,\,\, \dot{\vec{L}}=\vec n \times 
\vec{M}_{ext}+F_0\,\dot{\vec n}-\kappa_0\,(\vec n \times \ddot{\vec n})
\end{equation*}
where we have used standard properties of the vector triple product together 
with Eqs.~\eqref{conservation}. In the equations above we can substitute $\vec 
L\rightarrow \vec{L}_{\perp}$ where by definition $\vec{L}_\perp=\vec L -(\vec L 
\cdot \vec n)\vec n$. We now compute $I \ddot{\vec{n}}=\dot{\vec L}_\perp \times 
\vec n+\vec L_\perp \times \dot{\vec n}$ and using standard properties of the 
vector triple product together with $\vec{L}_\perp=I\,(\vec n \times \dot{\vec 
n})$ and Eqs.~\eqref{conservation} we arrive at:
\begin{equation}
I_{eff} \ddot{\vec{n}}=\left(\vec{n}\times\vec{M}_{ext}\right)\times \vec n + 
F_0 (\dot{\vec n} \times \vec n)-I_{eff} |\dot{\vec n}|^2 \vec n
\label{rotator}
\end{equation}
where we have defined the renormalized momentum of inertia is 
$I_{eff}=I+\kappa_0$.

From this equation we see that the moment of inertia of the rotator is 
renormalized by the interaction with the spin-$\frac{1}{2}$ particles.
Moreover we see that, even when the external force is absent 
($\vec{M}_{ext}=0$), the Berry curvature ($F_0$) causes the Coriolis type force 
tilting the rotation plane of the rotator. Indeed if we start with uniform 
rotations of the rotator in the $xz$ plane, i.e. $\vec n,\,\dot{\vec n}$ lie in 
the $xz$ plane, we immediately see that the Berry curvature causes acceleration 
orthogonal to the rotation plane. 
The physics behind the Coriolis force is intuitively simple. At any finite 
rotator's velocity, the spins will not be able to follow adiabatically the 
rotator and thus will be somewhat behind. As a result there will be a finite 
angle between the instantaneous direction of the spins and the rotator so the 
spins will start precessing around the rotator. This precession will result in a 
finite tilt of the spin orientation with respect to the rotation plane 
proportional to the Berry curvature~\cite{gritsev_2012}. In turn this tilt will 
result in precession of the rotator around the spins and cause the tilt of the 
rotation plane. It is interesting that despite the motion of the rotator is 
completely classical the Coriolis force given by the Berry curvature is quantum 
in nature. In particular, at zero temperature the integral of the Berry 
curvature over the closed manifold ($4\pi$ spherical angle is this case) is 
quantized. Therefore by measuring the Coriolis force and averaging it over the 
angles $\theta$ and $\phi$ one should be able to accurately see a quantized 
value:
\be
\int_0^\pi d\theta\int_0^{2\pi} d\phi\, F_{\theta,\phi}=2\pi N.
\ee
It is interesting that this result remains robust against any small 
perturbations in the system, which do not close the gap in the spectrum. 
Similarly the origin of the extra mass (moment of inertia) $\kappa$ in this 
simple example is purely quantum, i.e. despite this mass describes the classical 
Newtonian motion, it can not be computed within the classical framework.

\begin{figure}
\includegraphics[width=0.5\columnwidth]{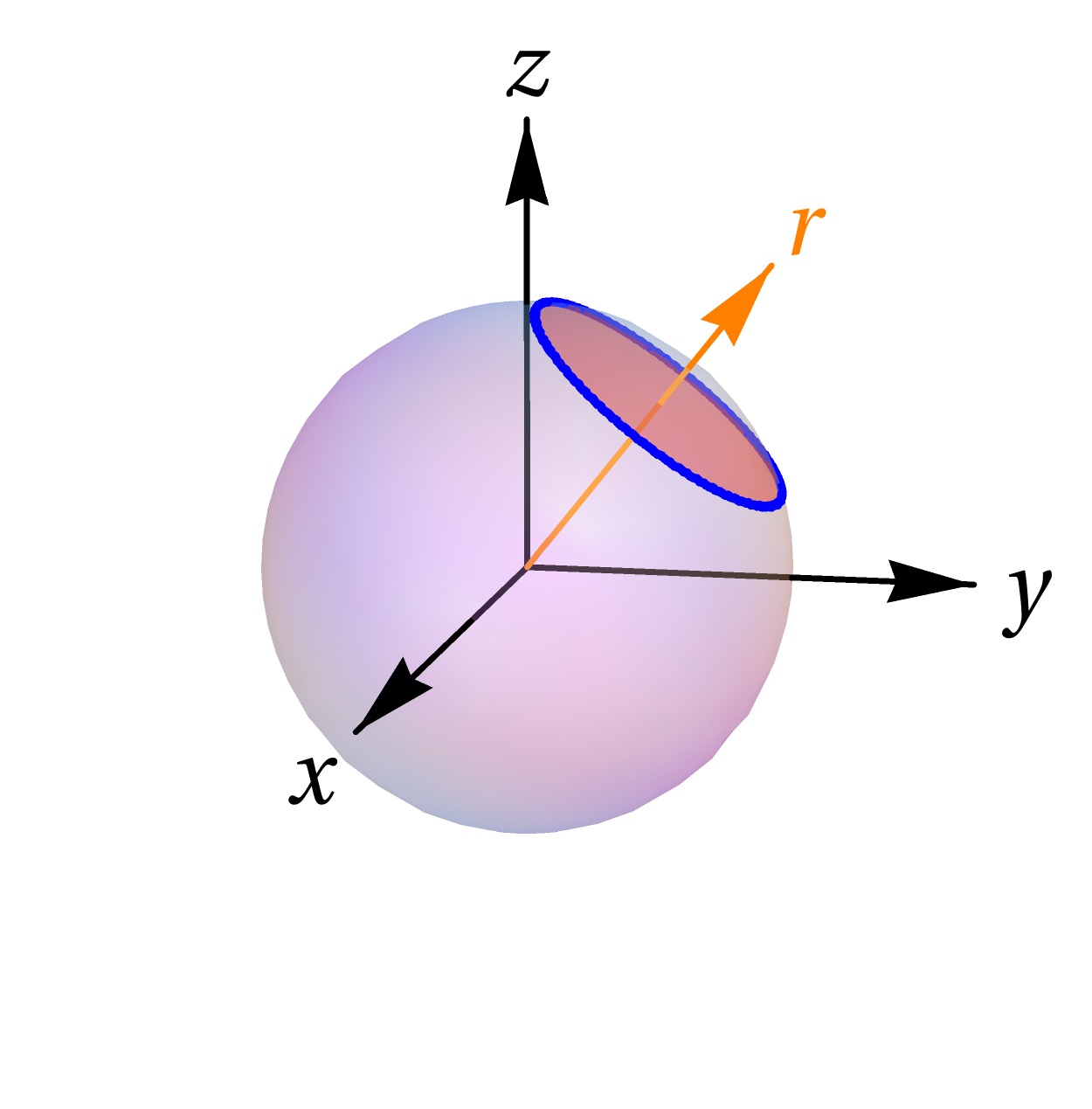}\includegraphics[
width=0.5\columnwidth]{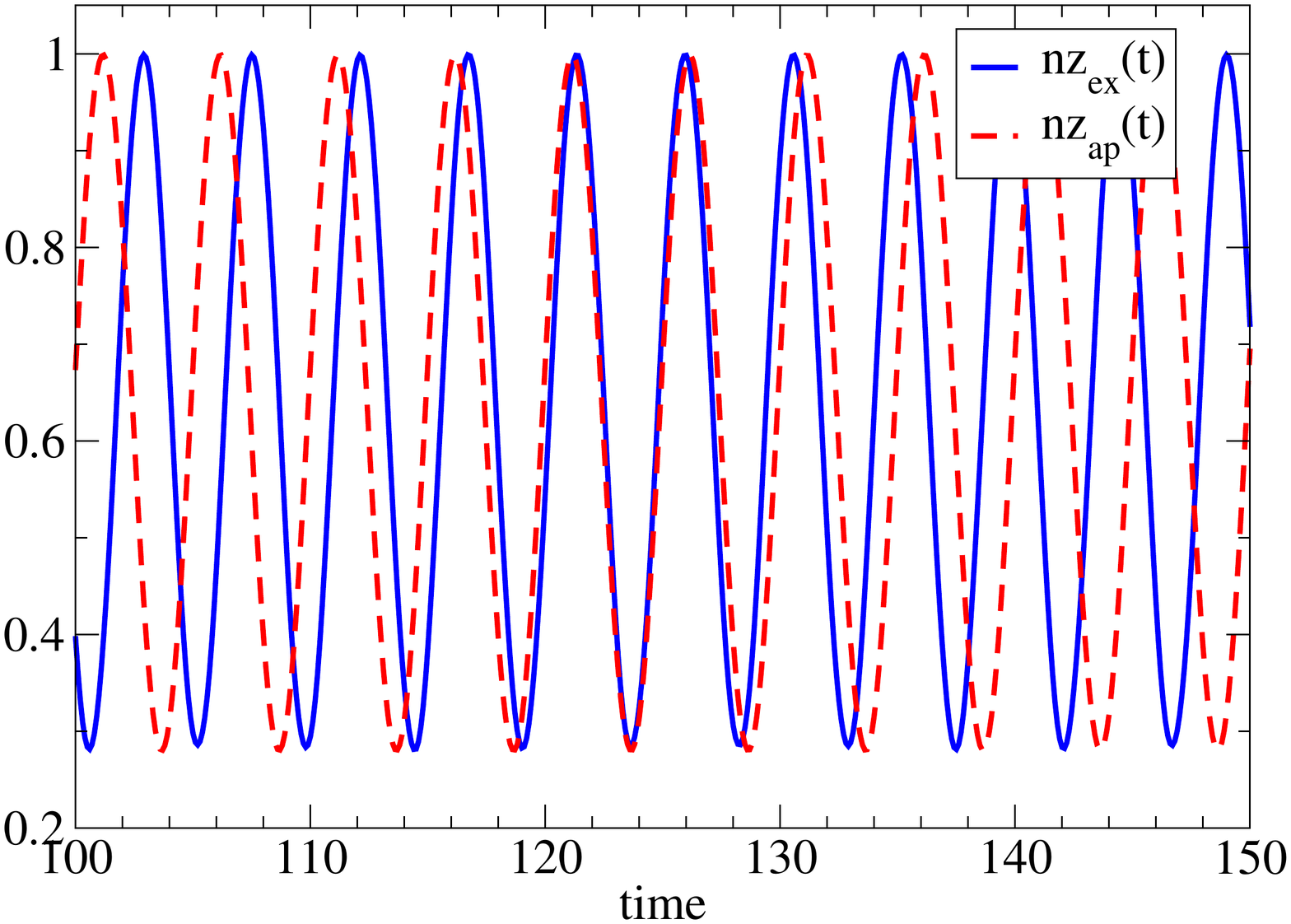}
\caption{(Color on-line) Dynamics of a rotator coupled to $N=20$ 
spins-$\frac{1}{2}$. (Left panel) The exact trajectory of the rotator obtained 
by numerically solving Eq.~\eqref{vonN} and \eqref{classical_rotor} (blue line) 
coincides with the approximated trajectory Eq.~\eqref{ansatz} which delimits the 
shaded red plane. (Right panel) Behavior of $n_z(t)$ for the exact (continuous 
blue) and approximated evolution (dashed red) versus time.  The parameters of 
the exact simulations are: $N=20$, $\beta=0.1$, $I=1$, $\Delta_i$ are randomly 
distributed in $(1,2)$. For these parameters $F_0\approx 1.5$, 
$\kappa_0\approx0.5$ and $I_{eff}/I\approx 1.5$. The initial conditions are 
$\vec{n}_0=(0,0,1)$ and $\vec{L}_0=(0,0,0)$ and initially the spins are in 
thermal equilibrium (see main text). We chose the time-dependent external force 
to be 
$\vec{M}_{ext}(t)=250\left[\frac{t}{t_c}\left(1-\frac{t}{t_c}\right)\right]^4\,
\hat{x}$ for $0\le t\le t_c=3$ and zero otherwise.}
\label{fig:Berry}
\end{figure}

We now analyze the approximated Equation~\eqref{rotator} in detail. 
We consider the situation in which $\vec{M}_{ext}(t)$ is slowly turned on (off) 
at time $t=0$ ($t=t_c$). 
When $\vec{M}_{ext}(t)=0$ Eq.~\eqref{rotator} describes a uniform circular 
motion whose solution can be written as:
\begin{equation}
\vec{n}_{ap}(t)=\sqrt{1-A_{ap}^2}\,\,\hat{r} + A_{ap}\,\,\left[ \cos(\omega_{ap} 
t+\phi)\,\,\hat{c}_1 +\sin(\omega_{ap} t+\phi)\,\,\hat{c}_2\right]
\label{ansatz}
\end{equation}
where $A_{ap}$ is the amplitude, $\omega_{ap}$ is the angular frequency, 
$\hat{r}$ is the vector orthogonal to the plane of motion (see 
Fig.~\ref{fig:Berry}) and $\hat{c}_1$, $\hat{c}_2$ are two orthogonal unit 
vectors spanning the plane of motion. 
Substituting the ansatz~\eqref{ansatz} in Eq.~\eqref{rotator} (with 
$\vec{M}_{ext}=0$) we obtain
\begin{equation}
\omega_{ap}=-\frac{F_0}{I_{eff}\sqrt{1-A_{ap}^2}}.
\label{prediction}
\end{equation}
The amplitude and the orientation of the plane of motion can not be computed 
from the initial conditions of Eq.~\eqref{rotator} since this equation is only 
valid after a transient time. We therefore extract them from the exact numerical 
solution:
\begin{equation*}
\sqrt{1-A_{ap}^2}\,\,\hat{r}=\lim_{T\rightarrow 
\infty}\frac{1}{T}\int_{t_c}^{t_c+T} dt'\,\vec{n}_{ex}(t')
\end{equation*}
and compute $\omega_{ap}$ via Eq.~\eqref{prediction}.
In Fig.~\ref{fig:Berry} we compare the exact trajectory $\vec{n}_{ex}(t)$ 
obtained by numerically solving Eq.~\eqref{vonN} and Eq.~\eqref{classical_rotor} 
for the rotator coupled to $N=20$ spins with gaps $\Delta_i$ uniformly 
distributed between $(1,2)$ to the approximated trajectory 
$\vec{n}_{ap}(t)$~\eqref{ansatz}. We note that the frequency of the approximated 
motion (estimated via Eq.\eqref{prediction}) underestimates the exact frequency 
by $8\%$ however if we had used the bare momentum of inertia in 
\eqref{prediction} with the chosen simulation parameters ($I_{eff}/I \approx 
1.5$) we would have overestimated the exact frequency by $28\%$. The accuracy 
will be higher if we increase the number of spins $N$. 

\section{Conclusions}

In this paper we showed how macroscopic Newtonian dynamics for slow degrees of 
freedom coupled to an arbitrary system emerges in the leading order of expansion 
about the adiabatic limit. Our results apply to open and closed systems, quantum 
and classical.   In particular, we found closed microscopic expressions for the 
friction tensor and for the mass  tensor due to dressing of the macroscopic 
parameter with excitations. We showed that the mass tensor is directly related 
to the changing geometry of the Hilbert space of the system due to the adiabatic 
motion of the degree of freedom. In this sense its origin is similar in nature 
to the origin of the dynamical Casimir 
force~\cite{dynamical_casimir,casimir_exp}. At a high temperature (or in the 
classical) limit the mass term becomes equal to the product of the inverse 
temperature and the Fubini-Study metric tensor describing the system. While we 
focused on coupling to ergodic systems, which equilibrate at finite temperatures 
in the absence of motion, our results are more general. In particular, they will 
apply to non-ergodic, e.g. integrable or glassy systems, as long as they reach 
some steady state (or approximately steady state). In this case the mass and the 
friction will be different from the equilibrium values. Thus the mass of the 
object coupled to a weakly non-integrable system can change in time as the 
latter slowly equilibrates.

Our results are non perturbative in the coupling between the macroscopic 
parameter and the (classical or quantum) system and can therefore be applied to 
situations in which mass renormalization of the macroscopic parameter is 
large~\cite{heavy_solitons}. This is in contrast with the standard 
many-body Green function approach \cite{mahan} in which the real and imaginary 
part of the self energy (which are responsible of the mass renormalization and 
dissipation respectively) are perturbative in the coupling. 
Moreover our integral equations ~\eqref{main} and \eqref{main_M} contain 
non-Markovian effects and are valid even when the relaxation time in the system 
is long. Our results are based on the adiabatic perturbation theory which is 
based on a separation of time-scales between the dynamics of the (slow) 
macroscopic parameter and the (fast) dynamics of the 
system~\cite{berry_robbins}. Therefore our approach is different and 
complementary to the usual Born-Markov scheme of open quantum 
systems~\cite{Lindblad,Markov,Petruccione}.

\section{Acknowledgments}
We gratefully acknowledge M. Kolodrubetz and E. Katz for many stimulating 
conversations and M. V. Berry for providing valuable feedback to this work and for pointing to us Ref.~\cite{berry_vector_potential}. 
This work was partially supported by BSF 2010318, NSF DMR-0907039, AFOSR 
FA9550-10- 1-0110

\appendix 
\renewcommand*{\thesection}{\Alph{section}}

\section{Appendix: Microscopic derivation of Eqs.~(\ref{main}) and 
(\ref{main_M})\label{sec:Derivation}}

Let us show the details of the derivation of Eq.~(\ref{main}) and 
Eq.~(\ref{main_M}). 
We start from the generic Hamiltonian $H(\lambda)$ with eigenstates 
$|m_\lambda\rangle$. 
Both the Hamiltonian and the eigenstates depend on time through the parameter 
$\lambda(t)$ which can be multicomponents.
For brevity we use the notation $H(t)\equiv H(\lambda(t))$ and 
$|m_t\rangle\equiv |m_{\lambda(t)}\rangle$. 
Our goal is to compute the evolution of the density matrix $\rho(t)$ subject to 
the 
initial condition $\left[\rho^0,H(\lambda_0)\right]=0$, i.e. the initial density 
matrix is stationary with respect the initial Hamiltonian.
In the course of our derivation we will implement a sequence of two unitary 
transformations.

First we make a unitary (time-dependent) transformation, $R(\lambda)$, from the 
fixed
(lab) frame to a new frame which is co-moving with the Hamiltonian. 
In the following we reserve the tilda sign to the quantities in the co-moving 
frame. 
In the co-moving frame the Hamiltonian $\tilde{H}(\lambda)\equiv 
R^\dagger(\lambda)H(\lambda)R(\lambda)$ assumes a diagonal form and its 
eigenstates $|\tilde{m}\rangle$ are $\lambda$-independent:
\be
\langle n_\lambda|H(\lambda)|m_\lambda\rangle=\langle 
\tilde{n}|\tilde{H}(\lambda)|\tilde{m}\rangle=\epsilon_{n}(\lambda)\,\delta_{n,m
}\label{spectrum}.
\ee
where $|m_\lambda\rangle\equiv R(\lambda) |\tilde{m}\rangle$. 
We note that $|\tilde{m}\rangle=|m_{\lambda_0}\rangle$ where 
$|m_{\lambda_0}\rangle$ is the static basis, which diagonalizes the Hamiltonian
at the initial value of $\lambda=\lambda_0$.  
In the co-moving frame the wave-function is 
$|\tilde{\psi}\rangle=R^{\dagger}(\lambda)|\psi\rangle$ and the density matrix 
is:
\[
\tilde{\rho}=R^\dagger(\lambda)\rho R(\lambda).
\]
We assume that the dynamical process starts at $t=0$ so that $R(0)$
is the identity operator and thus $\rho^{0}$ and $\tilde{\rho}^{0}$ coincide 
(this condition can be relaxed). 
The Hamiltonian which governs the time-evolution in the co-moving frame is 
$\tilde H_{eff}$:
\begin{equation}
\tilde{H}_{eff}\equiv\tilde{H}-\dot{\lambda}_{\nu}\tilde{\mathcal 
A}_{\nu}\label{eq:Heff}
\end{equation}
The first term $\tilde{H}$ is diagonal and it is therefore only responsible for 
phase
accumulation. The second ``Galilean'' type term (see e.g. 
Ref.~\cite{degrandi_2013})
originates from the fact that the basis transformation, $R(\lambda)$ is 
time-dependent
and plays the role of translation operator in the parameter 
space~\cite{kolodrubetz_2013}.
The operator 
\be
\tilde{\mathcal A}_{\nu}(\lambda)\equiv 
iR^\dagger(\lambda)\partial_{\lambda_{\nu}}R(\lambda)
\ee
is a gauge potential. Clearly it can be also written as $\mathcal 
A_\nu=i\partial_{\lambda_\nu}$ in a sense that
\be
i\langle n_\lambda|\overrightarrow{\partial}_{\lambda_\nu}| 
m_\lambda\rangle=\langle \tilde{n}| \tilde{\mathcal A}_\nu(\lambda)|\tilde{m} 
\rangle \label{AA}.
\ee
Gauge potentials are generally very complicated many-particle
operators. For systems coupled to a bath they involve both the system's
and bath's degrees of freedom as well as the system-bath interactions. In
the co-moving basis this term has off-diagonal components and is responsible for 
the
transition between the energy levels. 

Before proceeding let us illustrate this formalism with the example used in the 
main text (see Sec.~\ref{spin-example}).
The Hamiltonian in the fixed (lab) frame is: 
\begin{equation*}
H\left(\lambda\right)=-\Delta\left(\cos(\lambda)\,\sigma_{z}+\sin(\lambda)\,
\sigma_{x}\right)
\end{equation*}
with eigenstates 
\begin{equation*}
|gs_\lambda\rangle=\left(\begin{array}{c}
\cos\left(\frac{\lambda}{2}\right)\\
\sin\left(\frac{\lambda}{2}\right)
\end{array}\right),\quad|ex_\lambda\rangle=\left(\begin{array}{c}
-\sin\left(\frac{\lambda}{2}\right)\\
\cos\left(\frac{\lambda}{2}\right)
\end{array}\right)\label{eq:b}
\end{equation*}
The unitary transformation, which diagonalizes the instantaneous
Hamiltonian, $H(\lambda)$, is the rotation around the $y$-axis by the angle 
$\lambda$:
\begin{equation*}
R(\lambda)=\exp[-i \frac{\sigma_y}{2}\lambda]=\left(\begin{array}{cc}
\cos\left(\lambda/2\right) & -\sin\left(\lambda/2\right)\\
\sin\left(\lambda/2\right) & \cos\left(\lambda/2\right)
\end{array}\right)
\end{equation*}
Note that $R(\lambda_0=0)=\text{Identity}$. In the rotated (co-moving) frame we 
have:
\begin{equation*}
\tilde{H}\equiv R^\dagger(\lambda)H(\lambda)R(\lambda)=-\Delta\sigma_{z}
\end{equation*}
which is diagonal and (in this case) $\lambda$-independent. The eigenstates are 
trivially:
\begin{equation*}
|\tilde{gs}\rangle=\left(\begin{array}{c}
1\\
0
\end{array}\right),\quad|\tilde{ex}\rangle=\left(\begin{array}{c}
0\\
1
\end{array}\right)
\end{equation*}
which are identical to the eigenstates at $\lambda_0=0$: 
\begin{equation*}
|\tilde{gs}\rangle=|gs_{\lambda_0}\rangle,\,\,|\tilde{ex}\rangle=|ex_{\lambda_0}
)\rangle
\end{equation*}
Finally the Gauge potential is:
\begin{equation*}
\tilde{A}\equiv 
iR^{\dagger}(\lambda)\left(\partial_{\lambda}R(\lambda)\right)=\frac{\sigma_{y}}
{2},
\end{equation*}
which in this example is also $\lambda$-independent. Then Eq.~\eqref{AA} is 
verified by direct inspection.

Next, to derive Eq.~(\ref{main}) and Eq.~(\ref{main_M}), we solve the von 
Neumann's equation
\[
i\frac{d\tilde{\rho}}{dt}=\left[\tilde{H}-\dot{\lambda}_{\nu}\tilde{\mathcal{A}}
_{\nu},\tilde{\rho}\right]
\]
using standard time-dependent perturbation theory with the Galilean
term being the perturbation. The easiest way to do so is to go to the
interaction picture (the Heisenberg representation with respect to $\tilde{H}$), 
where the von Neumann's equation becomes 
\begin{equation}
i\,\frac{d\tilde{\rho}_{H}}{dt}=-\dot{\lambda}_{\nu}\left[\tilde{\mathcal{A}}_{H
,\nu}(t),\tilde{\rho}_{H}\right],\label{eq:rho}
\end{equation}
which is equivalent to the integral equation
\[
\tilde{\rho}_{H}(t)=\tilde{\rho}_{H}(0)+i\int_{0}^{t}dt'\dot{\lambda}_{\nu}
(t')\left[\tilde{\mathcal{A}}_{H,\nu}(t'),\tilde{\rho}_{H}(t')\right]
\]
Note that if the moving Hamiltonian $\tilde{H}$ is time dependent 
(which is the case when the energy spectrum depends explicitly on $\lambda$, see 
Eq.~\eqref{spectrum}) 
the Heisenberg picture is different from the conventional one. But
because $\tilde{H}$ is always diagonal in the co-moving basis the
difference appears only in the phase factor:
\begin{equation}
\begin{split}
\langle\tilde{n}|\tilde{O}_{H}(\lambda(t))|\tilde{m}\rangle 
&=\exp\left[i\int_{0}^{t}dt'(\epsilon_{n}(t')-\epsilon_{m}(t'))\right]\langle 
\tilde{n}|\tilde{O}(\lambda(t))|\tilde{m}\rangle\\
&=\exp\left[i\int_{0}^{t}dt'(\epsilon_{n}(t')-\epsilon_{m}(t'))\right]\langle 
n_t|O(\lambda(t))|m_t\rangle
\label{eq:Heisenberg}
\end{split}
\end{equation}
which follows from the definition
\begin{equation*}
\tilde{O}_{H}(t)\equiv \left(e^{i\int_0^t\,d\tau \tilde{H}(\tau)}\right) 
R^{\dagger}(t)O(t)R(t)\left(e^{-i\int_0^t\,d\tau \tilde{H}(\tau)}\right)
\end{equation*}
We emphasize that this expression is not the same as the Heisenberg 
representation with respect to the original Hamiltonian $H(\lambda(t))$. The 
representation we use is perhaps more correctly termed as the adiabatic 
Heisenberg representation since it uses adiabatic energy levels $\epsilon_n(t)$, 
while all the transitions (off-diagonal terms) are treated perturbatively. 
Clearly if $\epsilon_{n}$ is time independent the 
expression~\eqref{eq:Heisenberg}
reduces to the conventional Heisenberg representation of the operator 
$O(\lambda(t))$ which can depend on time 
through the parameter $\lambda(t)$.
We can now solve Eq.~(\ref{eq:rho}) iteratively to the second order
in $\dot{\lambda}$: 
\begin{equation*}
\begin{split}
\partial_{t}\tilde{\rho}_{H}(t)&=i\dot{\lambda}_{\nu}(t)\left[\tilde{\mathcal{A}
}_{H,\nu}(t),\rho^{0}\right]\\
&-\dot{\lambda}_{\nu}(t)\int_{0}^{t}dt'\dot{\lambda}_{\mu}(t')\left[\tilde{
\mathcal{A}}_{H,\nu}(t),\left[\tilde{\mathcal{A}}_{H,\mu}(t'),\rho^{0}\right]
\right]+\mathcal{O}(|\dot{\lambda}|^{3}),
\end{split}
\end{equation*}
where we used $\tilde{\rho}_{H}^{0}=\tilde{\rho}^{0}=\rho^{0}$.

To derive Eq.~\eqref{main} we analyze the energy generation rate
\[
\partial_{t}\langle H\rangle=\partial_{t}\left( {\rm 
Tr}\left\{\tilde{\rho}_H(t)\,\tilde{H}(t)\right\}\right)=\dot{W}(t)+\dot{Q}(t),
\]
where 
\[
\dot{W}(t)={\rm Tr}\left\{ \tilde{\rho}_{H}(t)\,(\partial_{t}\tilde{H})\right\} 
,\;\dot{Q}(t)={\rm Tr}\left\{ 
(\partial_{t}\tilde{\rho}_{H})\,\tilde{H}(t)\right\} 
\]
The adiabatic work rate, related to the change of the spectrum in the 
Hamiltonian,
in turn can be rewritten as 
\[
\dot{W}(t)=-\dot{\lambda}_{\nu}M_{\nu}+\mathcal{O}\left(|\dot{\lambda}|^{3}
\right),
\]
 where 
\[
M_{\nu}=-\sum_{n}\langle 
n_0|\rho^{0}|n_0\rangle\left(\partial_{\lambda_{\nu}}\epsilon_{n}
(\lambda)\right)\equiv-\langle\partial_{\lambda_{\nu}}H\rangle_{0}\equiv\langle 
\mathcal{M}_{\nu}\rangle_{0}
\]
is the generalized force with respect the initial (stationary) density
matrix and we have defined $\mathcal{M}_{\nu}\equiv-\partial_{\lambda_{\nu}}H$.
For an initially stationary density matrix, $\left[ \rho^0, 
H(\lambda_0)\right]=0$, it is easy to see that
$\left[ \tilde{\rho}^0,\partial_t \tilde{H}\right]=0$ (recall that 
$\tilde{H}(t)$ is diagonal for any time $t$)
and there is no second order contribution in $\dot{\lambda}$ to the work rate. 
In general we find that for an initially stationary
density matrix the work rate (heat rate) is odd (even) in $\dot{\lambda}$.
Then the leading contribution to the heat rate $\dot{Q}$ , related
to the change of occupation of the instantaneous eigenstates of the
Hamiltonian, is quadratic in $\dot\lambda$ and we find
\begin{equation}
\begin{split}
\dot{Q}(t)&=-\dot{\lambda}_{\nu}(t)\int_{0}^{t}dt'\dot{\lambda}_{\mu}(t')\times 
{\rm Tr}\left\{ 
\tilde{H}(t)\left[\tilde{\mathcal{A}}_{H,\nu}(t),\left[\tilde{\mathcal{A}}_{H,
\mu}(t'),\rho^{0}\right]\right]\right\} \\
&=\dot{\lambda}_{\nu}(t)\int_{0}^{t}dt'\dot{\lambda}_{\mu}(t')\,\,\left<\left[
\left[\tilde{\mathcal{A}}_{H,\nu}(t),\tilde{H}(t)\right],\tilde{\mathcal{A}}_{H,
\mu}(t')\right]\right>_{0}
\label{eq:formally-exact}
\end{split}
\end{equation}
Note that in this expression the Hamiltonian $\tilde{H}(t)$ depends on time only
through the spectrum and it is diagonal for any time $t$. 
From the definition of $\tilde{\mathcal{A}}_{\nu}\equiv 
iR^\dagger\left(\partial_{\lambda_{\nu}}R\right)$
and $\tilde{H}\equiv R^\dagger H R$ it follows that:
\begin{equation}
\left[\tilde{\mathcal{A}}_{\nu}(t),\tilde{H}(t)\right]=-i\tilde{\mathcal 
M}_{\nu}-i\partial_{\lambda_{\nu}}\tilde{H}
\label{eq:commutator}
\end{equation}
where 
\[
\tilde{\mathcal M}_{\nu}\equiv R^\dagger \mathcal M_{\nu}R \equiv 
-R^\dagger\left(\partial_{\lambda_{\nu}}H\right) R
\]
is the generalized force in the co-moving basis and we have used 
the identity 
$\left(\partial_{\lambda_{\nu}}R\right)R^\dagger=-R\left(\partial_{\lambda_{\nu}
}R^\dagger\right)$.
Eq.~(\ref{eq:commutator}) naturally follows from the
interpretation of the gauge potential $\mathcal{A}_{\nu}$ as the (negative) 
momentum
operator along the direction $\lambda_{\nu}$: 
$\mathcal{A}_{\nu}=i\partial_{\lambda_{\nu}}$ (see Eq.~\eqref{AA}).
The second term in the RHS of Eq.~(\ref{eq:commutator}) is diagonal
in the co-moving basis and does not contribute to Eq.~(\ref{eq:formally-exact}),
which then simplifies to 
\begin{equation}
\dot{Q}=-i\dot{\lambda}_{\nu}(t)\int_{0}^{t}dt'\dot{\lambda}_{\mu}(t')\,\,
\left<\left[\tilde{\mathcal 
M}_{H,\nu}(t),\tilde{\mathcal{A}}_{H,\mu}(t')\right]\right>_{0} 
\end{equation}
And finally, evaluating this expression in the co-moving basis (see 
Eq.~\eqref{eq:Heisenberg}), and using the identity 
\begin{equation}
\langle \tilde{n}|\tilde{\mathcal{A}}_{\mu}(\lambda)|\tilde{m}\rangle=i\langle 
n_\lambda|\partial_{\lambda_{\mu}}|m_\lambda\rangle =-i\frac{\langle 
n_\lambda|\mathcal 
M_{\mu}(\lambda)|m_\lambda\rangle}{\epsilon_{m}(\lambda)-\epsilon_{n}(\lambda)}
=-i\frac{\langle \tilde{n}|\tilde{\mathcal 
M}_{\mu}(\lambda)|\tilde{m}\rangle}{\epsilon_{m}(\lambda)-\epsilon_{n}(\lambda)}
\label{rel_AO}
\end{equation}
 we arrive at 
\begin{align}
&\dot{Q}(t)=\dot{\lambda}_{\nu}(t)\int_{0}^{t}dt'\,\dot{\lambda}_{\mu}
(t')\nonumber\\
&\times \sum_{n\neq 
m}\,\frac{\rho_{n}^{0}-\rho_{m}^{0}}{\epsilon_{m}(t')-\epsilon_{n}(t')} \mathrm 
e^{i\int_{t'}^t d\tau (\epsilon_m(\tau)-\epsilon_n(\tau))}\langle m_t|\mathcal 
M_{\nu}(\lambda(t))|n_t\rangle\langle n_{t'}|\mathcal 
M_{\mu}(\lambda(t'))|m_{t'}\rangle\nonumber \\
&=\dot{\lambda}_{\nu}(t)\int_{0}^{t}dt'\,\dot{\lambda}_{\mu}(t')\nonumber\\
&\times \sum_{n\neq 
m}\,\frac{\rho_{n}^{0}-\rho_{m}^{0}}{\epsilon_{m}(t')-\epsilon_{n}(t')} \mathrm 
e^{i\int_{t'}^t d\tau (\epsilon_m(\tau)-\epsilon_n(\tau))} \langle 
\tilde{m}|\tilde{\mathcal M}_{\nu}(\lambda(t))|\tilde{n}\rangle\langle 
\tilde{n}|\tilde{\mathcal M}_{\mu}(\lambda(t'))|\tilde{m}\rangle\nonumber \\
&=\dot{\lambda}_{\nu}(t)\int_{0}^{t}dt'\,\dot{\lambda}_{\mu}(t') \sum_{n\neq 
m}\,\frac{\rho_{n}^{0}-\rho_{m}^{0}}{\epsilon_{m}(t')-\epsilon_{n}(t')} \langle 
\tilde{m}|\tilde{\mathcal M}_{H,\nu}(t)|\tilde{n}\rangle\langle 
\tilde{n}|\tilde{\mathcal M}_{H,\mu}(t')|\tilde{m}\rangle\nonumber \\
&=\dot{\lambda}_{\nu}(t)\int_{0}^{t}dt'\,\int_0^\beta\,d\tau\,\,\dot{\lambda}_{
\mu}(t') \sum_{n\neq m}\,\rho_{m}^{0} \langle \tilde{m}|\tilde{\mathcal 
M}_{H,\nu}(t)|\tilde{n}\rangle\langle \tilde{n}|\tilde{\mathcal 
M}_{H,\mu}(t'+i\tau)|\tilde{m}\rangle \nonumber\\
&=\dot{\lambda}_{\nu}(t)\int_{0}^{t}dt'\,\int_0^\beta\,d\tau\,\,\dot{\lambda}_{
\mu}(t')\,\,\langle \tilde{\mathcal M}_{H,\nu}(t) \tilde{\mathcal 
M}_{H,\mu}(t'+i\tau)\rangle_0,
\label{main_supp}
\end{align}
Eq.~\eqref{main_supp} gives the microscopic heat production rate in the most 
general form and in the last two lines we have used
the fact that, for thermal distribution, $\rho^0_m=Z^{-1}\exp[-\beta 
\epsilon_m]$, we have 
\begin{equation}
\int_0^\beta\,d\tau\,\,\rho^0_m\,\,e^{-(\epsilon_n-\epsilon_m)\tau}=\frac{
\rho^0_n-\rho^0_m}{\epsilon_m-\epsilon_n}
\label{im_int}
\end{equation}

The derivation of Eq.~(\ref{main_M}) is even simpler since it requires only 
going to the first order perturbation theory:
\be
\tilde{\rho}_{H}(t)\approx \rho_0+i\int_0^t dt'\,\dot\lambda_\mu(t') 
\left[\tilde{\mathcal A}_{H,\mu}(t'),\rho_0\right].
\ee
Therefore
\begin{equation*}
\langle \mathcal M_\nu(t)\rangle=M_\nu(t)+i\int_0^t dt'\,\dot\lambda_\mu(t') 
\langle [\tilde{\mathcal M}_{H,\nu}(t),\tilde{\mathcal A}_{H,\mu}(t')]\rangle_0,
\end{equation*}
where we recall that by definitions $M_\nu(t)\equiv\langle \mathcal 
M_\nu(t)\rangle_0$.
Evaluating this expression in the co-moving basis (see 
Eq.~\eqref{eq:Heisenberg}) and using the identity~\eqref{rel_AO} we arrive at
\begin{align}
&\langle \mathcal M_\nu(t)\rangle=M_\nu(t)-\int_0^t 
dt'\,\dot\lambda_\mu(t')\nonumber\\
&\times \sum_{n\ne m} 
\,\frac{\rho_n^0-\rho_m^0}{\epsilon_m(t')-\epsilon_n(t')}\mathrm e^{i\int_{t'}^t 
d\tau (\epsilon_m(\tau)-\epsilon_n(\tau))} \langle m_t|\mathcal 
M_{\nu}(\lambda(t))|n_t\rangle \langle n_{t'}|\mathcal 
M_{\mu}(\lambda(t'))|m_{t'}\rangle\nonumber \\
&=M_\nu(t)-\int_0^t dt'\,\dot\lambda_\mu(t')\nonumber\\
&\times \sum_{n\ne m} 
\,\frac{\rho_n^0-\rho_m^0}{\epsilon_m(t')-\epsilon_n(t')}\mathrm e^{i\int_{t'}^t 
d\tau (\epsilon_m(\tau)-\epsilon_n(\tau))} \langle \tilde{m}|\tilde{\mathcal 
M}_{\nu}(\lambda(t))|\tilde{n}\rangle \langle \tilde{n}|\tilde{\mathcal 
M}_{\mu}(\lambda(t'))|\tilde{m}\rangle \nonumber\\
&=M_\nu(t)-\int_0^t dt'\,\dot\lambda_\mu(t') \sum_{n\ne m} 
\,\frac{\rho_n^0-\rho_m^0}{\epsilon_m(t')-\epsilon_n(t')}\langle 
\tilde{m}|\tilde{\mathcal M}_{H,\nu}(t)|\tilde{n}\rangle \langle 
\tilde{n}|\tilde{\mathcal M}_{H,\mu}(t')|\tilde{m}\rangle\nonumber \\
&=M_\nu(t)-\int_0^t dt'\int_0^\beta\,d\tau\,\,\dot\lambda_\mu(t') \sum_{n\ne m} 
\,\rho_m^0 \langle \tilde{m}|\tilde{\mathcal M}_{H,\nu}(t)|\tilde{n}\rangle 
\langle \tilde{n}|\tilde{\mathcal M}_{H,\mu}(t'+i\tau)|\tilde{m}\rangle\nonumber 
\\
&=M_\nu(t)-\int_0^t dt'\int_0^\beta d\tau\,\,\dot\lambda_\mu(t') \langle 
\tilde{\mathcal M}_{H,\nu}(t)\tilde{\mathcal M}_{\mu}(t'+i\tau)\rangle_0,
\label{M_supp}
\end{align}
which gives the microscopic force in the most general form.

In Eqs.~\eqref{main_supp} and \eqref{M_supp} we now perform a change of dummy 
integration variable $t'\rightarrow t-t'$ and observe that the leading order in 
$|\dot{\lambda}|$ corresponds to evaluate the spectrum, the eigenstates and the 
force at the {\it final} time of the evolution, i.e. at $t'=0$. 
This is because the energies, the forces and the eigenstates depend on time only 
through the parameter $\lambda(t)$ and can therefore be expanded in powers of 
$|\dot{\lambda}|$. For example
\begin{equation*}
\epsilon_m(t-t')\equiv\epsilon_m(\lambda(t-t'))=\epsilon_m(t)-t'\dot{\lambda}
(t)\partial_{\lambda}\epsilon_m+\dots \approx \epsilon_m(t)
\end{equation*}
We then obtain the leading contributions:
\begin{align}
&\dot{Q}(t)=\dot{\lambda}_{\nu}(t)\int_{0}^{t}dt'\,\dot{\lambda}_{\mu}
(t-t')\nonumber \\
&\times \sum_{n\neq m}\,\,e^{i(\epsilon_m(t)-\epsilon_n(t))t'} 
\frac{\rho_{n}^{0}-\rho_{m}^{0}}{\epsilon_{m}(t)-\epsilon_{n}(t)} \langle 
m_t|\mathcal M_{\nu}(\lambda(t))|n_t\rangle\langle n_t|\mathcal 
M_{\mu}(\lambda(t))|m_t\rangle \\
&\langle \mathcal M_\nu(t)\rangle=M_\nu(t)-\int_0^t 
dt'\,\dot\lambda_\mu(t-t')\nonumber \\
&\times \sum_{n\ne m} e^{i 
(\epsilon_m(t)-\epsilon_n(t))t'}\frac{\rho_n^0-\rho_m^0}{
\epsilon_m(t)-\epsilon_n(t)} \langle m_t|\mathcal M_{\nu}(\lambda(t))|n_t\rangle 
\langle n_t|\mathcal M_{\mu}(\lambda(t))|m_t\rangle
\label{Lehmann_supp}
\end{align}
which for time independent spectrum, $\epsilon_{m}(\lambda(t))=\epsilon_{m}$, 
and time independent force, $\mathcal M_{\mu}(\lambda(t))=\mathcal M_{\mu}$ can 
be rewritten as the equations in the main text (see Eqs.~\eqref{main} and 
\eqref{main_M}):
\begin{equation}
\begin{split}
\dot{Q}(t)&=\dot{\lambda}_{\nu}(t)\int_{0}^{t}dt'\,\int_0^{\beta} 
\dot{\lambda}_{\mu}(t-t') \langle \mathcal M_{H,\nu}(t')\mathcal 
M_{H,\mu}(i\tau)\rangle_0 \\
\langle \mathcal M_\nu\rangle&=M_\nu-\int_{0}^{t}dt'\,\int_0^{\beta} 
\dot{\lambda}_{\mu}(t-t') \langle \mathcal M_{H,\nu}(t')\mathcal 
M_{H,\mu}(i\tau)\rangle_0 
\end{split}
\label{final_result}
\end{equation}
where we have used the fact that, for thermal distribution, 
$\rho^0_m=Z^{-1}\exp[-\beta \epsilon_m]$, we have 
\begin{equation*}
\int_0^\beta\,d\tau\,\,\rho^0_m\,\,e^{-(\epsilon_n-\epsilon_m)\tau}=\frac{
\rho^0_n-\rho^0_m}{\epsilon_m-\epsilon_n}
\end{equation*}
The expressions~\eqref{final_result} are correct even if the spectrum and force 
are time dependent but vary slowly on the scale of the relaxation time in the 
system.

\section{Appendix: General structure of expansion of Eq.~(\ref{main}) and 
Eq.~(\ref{main_M}) in time derivatives of 
$\vec{\lambda}$\label{sec:time-derivatives}}

Here we analyze the full Taylor expansion of Eq.~(\ref{main}) and and 
Eq.~(\ref{main_M}), which
we rewrite in the Lehmann's representation (see Eq.~\eqref{Lehmann_supp}).
In order for the following expansion to be valid both the velocity and the 
energy spectrum need to change slowly compared to the relaxation time scale on 
the system, i.e. the relaxation time must be the shortest time scale.
For times $t$ longer than the relaxation time we can set the upper limit of 
integration in Eq.~\eqref{Lehmann_supp} to infinity, i.e. $t\to\infty$. 
Expanding
$\dot{\lambda_\mu}(t-t')$ into the Taylor series 
\[
\dot{\lambda_\mu}(t-t')=\sum_{k=1}^{\infty}\lambda_{\mu}^{(k)}(t)\frac{(-t')^{
k-1}}{(k-1)!}
\]
we find:
\begin{equation}
\begin{split}
\dot{Q}(t)&=\dot{\lambda}_{\nu}(t)\left(\sum_{k=1}^{\infty}\, 
\chi_{\nu,\mu}^k(t)\, \lambda_{\mu}^{(k)}(t)\right) \\
\langle \mathcal M_\nu(t)\rangle&=M_\nu(t)-\sum_{k=1}^{\infty}\, 
\chi_{\nu,\mu}^k(t) \, \lambda_{\mu}^{(k)}(t)
\end{split}
\end{equation}
where we have defined:
\begin{multline}
\chi_{\nu,\mu}^k(t)\equiv \int_0^\infty dt' \,\frac{(-t')^{k-1}}{(k-1)!} 
\sum_{n\neq m} \mathrm{e}^{i(\epsilon_{m}(t)-\epsilon_{n}(t))t'-|\varepsilon|t'} 
\times \\ 
\frac{\rho_{n}^{0}-\rho_{m}^{0}}{\epsilon_{m}(t)-\epsilon_{n}(t)}  \,\langle 
m_t|\mathcal{M}_{\nu}(\lambda(t))|n_t\rangle \langle 
n_t|\mathcal{M}_{\mu}(\lambda(t))|m_t\rangle
\end{multline}
and we have added an infinitesimal imaginary energy $|\varepsilon|$ for 
convergence. 
Performing time integration over $t'$ we immediately find:
\begin{multline}
\chi_{\nu,\mu}^k(t) = \sum_{n\neq m} 
\frac{(-1)^k}{\left(|\varepsilon|+i(\epsilon_{n}(t)-\epsilon_{m}(t))\right)^k}\,
\times\\ 
\frac{\rho_{n}^{0}-\rho_{m}^{0}}{\epsilon_{n}(t)-\epsilon_{m}(t)} \langle 
m_t|\mathcal{M}_{\nu}(\lambda(t))|n_t\rangle \langle 
n_t|\mathcal{M}_{\mu}(\lambda(t))|m_t\rangle
\label{LehmanK}
\end{multline}
Using the identities
\begin{equation}
\frac{1}{|\varepsilon|+i x}=\pi \delta(x) - \frac{i}{x},\quad 
\frac{1}{(|\varepsilon|+i x)^2}=i \pi \delta'(x) - \frac{1}{x^2} 
\label{identities}
\end{equation}
eq.~\eqref{LehmanK} can be used to write explicitly the Lehmann representation 
of the tensors $\eta$, $\kappa$, $F$ and $F'$:
\begin{equation*}
\begin{split}
&\eta_{\nu,\mu}(t)\equiv \frac{\chi_{\nu,\mu}^1(t)+\chi_{\mu,\nu}^1(t)}{2},\quad 
F_{\nu,\mu}(t)\equiv -\frac{\chi_{\nu,\mu}^1(t)-\chi_{\mu,\nu}^1(t)}{2} \\
&\kappa_{\nu,\mu}(t)\equiv 
\frac{\chi_{\nu,\mu}^2(t)+\chi_{\mu,\nu}^2(t)}{2},\quad 
F^{\prime}_{\nu,\mu}(t)\equiv \frac{\chi_{\nu,\mu}^2(t)-\chi_{\mu,\nu}^2(t)}{2} 
\end{split}
\end{equation*} 
The extra minus sign in the definition of $F_{\nu,\mu}$ is conventional and it 
is chosen to reproduce known expression for the Berry curvature.

To shorten the notation we now drop the time label but in all the following 
expressions the energy, the force and the eigenstates should be understood as 
(slowly) depending on time through the parameter $\lambda(t)$.
Equation~\eqref{LehmanK} can be rewritten as
\begin{equation*}
\chi_{\nu,\mu}^k = \int_{-\infty}^{\infty}d\omega\,\sum_{n\neq 
m}\,\delta(\epsilon_n-\epsilon_m-\omega) 
\frac{(-1)^k}{\left(|\varepsilon|+i\omega)\right)^k} 
\frac{\rho_{n}^{0}-\rho_{m}^{0}}{\omega} \langle m|\mathcal{M}_{\nu}|n\rangle 
\langle n|\mathcal{M}_{\mu}|m\rangle.
\end{equation*} 
Expressing 
\begin{equation*}
\delta(\epsilon_n-\epsilon_m-\omega)=\int_{-\infty}^{\infty}\frac{dt}{2\pi}\,
\mathrm{e}^{-i(\epsilon_n-\epsilon_m-\omega)t}
\end{equation*}
we obtain the compact expression:
\be
\chi_{\nu,\mu}^k = -\int_{-\infty}^{\infty}d\omega \, 
\frac{(-1)^k}{\left(|\varepsilon|+i\omega)\right)^k}\,\left(\frac{\mathcal{S}_{
\nu,\mu}(\omega)}{\omega}\right) 
\label{GRK}
\ee
where we have defined the spectral density $\mathcal{S}_{\nu,\mu}(t)$ and its 
Fourier Transform:
\begin{equation*}
\begin{split} 
\mathcal{S}_{\nu,\mu}(t) &\equiv 
\frac{1}{2\pi}\,\langle\left[\mathcal{M}_{H,\nu}(t),\mathcal{M}_{H,\mu}(0)\right
]\rangle_0 \\
\mathcal{S}_{\nu,\mu}(\omega) &\equiv \int_{-\infty}^{\infty}dt\,\mathrm{e}^{i 
\omega t}\, \mathcal{S}_{\nu,\mu}(t) 
\end{split}
\end{equation*} 
From the Lehmann representation
\begin{equation*}
\mathcal{S}_{\nu,\mu}(\omega)=-\sum_{n\ne m} 
\delta(\epsilon_n-\epsilon_m-\omega) (\rho_n^0-\rho_m^0) \langle 
m|\mathcal{M}_{\nu}|n\rangle \langle n|\mathcal{M}_{\mu}|m\rangle
\end{equation*}
it is clear that 
\be
\mathcal{S}_{\nu,\mu}(\omega)=-\mathcal{S}_{\mu,\nu}(-\omega)
\label{sym_S}
\ee
Moreover the fluctuation-dissipation relation can be written as
\begin{equation*}
\Phi_{\nu,\mu}(\omega)=\coth\left(\frac{\beta \omega}{2}\right) 
\mathcal{S}_{\nu,\mu}(\omega)
\end{equation*}
where we have defined  
\begin{equation*}
\begin{split}
\Phi_{\nu,\mu}(\omega)&\equiv\frac{1}{2\pi} 
\int_{-\infty}^{\infty}dt\,\mathrm{e}^{i \omega t}\, 
\langle\left\{\mathcal{M}_{H,\nu}(t),\mathcal{M}_{H,\mu}(0)\right\}\rangle_0 \\
&=\sum_{n\ne m} \delta(\epsilon_n-\epsilon_m-\omega) (\rho_n^0+\rho_m^0) \langle 
m|\mathcal{M}_{\nu}|n\rangle \langle n|\mathcal{M}_{\mu}|m\rangle
\end{split}
\end{equation*}
which satisfy the symmetry relation
\begin{equation*}
\Phi_{\nu,\mu}(\omega)=\Phi_{\mu,\nu}(-\omega).
\end{equation*}

Finally using the identities \eqref{identities} we obtain:
\begin{equation*}
\begin{split}
\chi_{\nu,\mu}^1 &= \pi \sum_{n\neq m}\,\frac{\rho_n^0-\rho_m^0}{\epsilon_m-\epsilon_n} \langle m|\mathcal{M}_{\nu}|n\rangle \langle n|\mathcal{M}_{\mu}|m\rangle\,\delta(\epsilon_n-\epsilon_m) \\
&+i \sum_{n\neq m}\,\frac{\rho_n^0-\rho_m^0}{(\epsilon_n-\epsilon_m)^2} \langle m|\mathcal{M}_{\nu}|n\rangle \langle n|\mathcal{M}_{\mu}|m\rangle \\
&=\pi \mathcal{G}_{\nu,\mu}(\omega=0)-i \mathcal{P}\int_{-\infty}^{+\infty} 
d\omega\,\frac{\mathcal{S}_{\nu,\mu}(\omega)}{\omega^2}\\
&= \frac{\pi}{2} 
\left(\mathcal{G}_{\nu,\mu}(\omega=0)+\mathcal{G}_{\mu,\nu}
(\omega=0)\right)-\frac{i}{2} \mathcal{P}\int_{-\infty}^{+\infty} 
\frac{d\omega}{\omega^2}\,\left(\mathcal{S}_{\nu,\mu}(\omega)-\mathcal{S}_{\mu,
\nu}(\omega)\right)\\
\chi_{\nu,\mu}^2 &= i \pi \sum_{n\neq m}\,\frac{\rho_n^0-\rho_m^0}{\epsilon_n-\epsilon_m} \langle m|\mathcal{M}_{\nu}|n\rangle \langle n|\mathcal{M}_{\mu}|m\rangle\,\delta'(\epsilon_n-\epsilon_m) \\
&+ \sum_{n\neq m}\,\frac{\rho_n^0-\rho_m^0}{(\epsilon_m-\epsilon_n)^3} \langle 
m|\mathcal{M}_{\nu}|n\rangle \langle n|\mathcal{M}_{\mu}|m\rangle \\
&=i \pi \partial_{\omega}\mathcal{G}_{\nu,\mu}|_{\omega=0}+ 
\mathcal{P}\int_{-\infty}^{+\infty} 
d\omega\,\frac{\mathcal{S}_{\nu,\mu}(\omega)}{\omega^3}\\
&=\frac{i \pi}{2} \left( 
\partial_{\omega}\mathcal{G}_{\nu,\mu}|_{\omega=0}-\partial_{\omega}\mathcal{G}_
{\mu,\nu}|_{\omega=0}\right)+ \frac{1}{2}\mathcal{P}\int_{-\infty}^{+\infty} 
\frac{d\omega}{\omega^3}\,\left(\mathcal{S}_{\nu,\mu}(\omega)+\mathcal{S}_{\mu,
\nu}(\omega)\right) 
\end{split}
\end{equation*}
where $\mathcal{P}$ indicates the principal value, used the symmetry of 
$\mathcal{S}$ (see Eq.~\eqref{sym_S}) and we have defined the auxiliary function 
$\mathcal{G}_{\nu,\mu}(\omega)\equiv\frac{\mathcal{S}_{\nu,\mu}(\omega)}{\omega}
=\mathcal{G}_{\mu,\nu}(-\omega)$.

We note that:
\begin{equation*}
F_{\nu,\mu}\equiv i \langle \overleftarrow{\partial}_{\lambda_\nu} 
\overrightarrow{\partial}_{\lambda_\mu}-\overleftarrow{\partial}_{\lambda_\mu} 
\overrightarrow{\partial}_{\lambda_\nu}\rangle_0
\end{equation*}
and that when the occupations are thermal we can write:
\begin{equation*}
\begin{split}
&\eta_{\nu,\mu}= \pi \beta \sum_{n\neq m}\,\rho_m^0 \langle 
m|\mathcal{M}_{\nu}|n\rangle \langle 
n|\mathcal{M}_{\mu}|m\rangle\,\delta(\epsilon_n-\epsilon_m) \\
&F^{\prime}=-i \pi \sum_{n\neq m}\,\rho_m^0\,\left[ \beta^2 \langle 
m|\mathcal{M}_{\nu}|n\rangle \langle n|\mathcal{M}_{\mu}|m\rangle \right. \\
&-\left.\beta \frac{\partial}{\partial(\epsilon_n-\epsilon_m)} \left(\langle 
m|\mathcal{M}_{\nu}|n\rangle \langle 
n|\mathcal{M}_{\mu}|m\rangle\right)\right]\,\delta(\epsilon_n-\epsilon_m). 
\end{split}
\end{equation*}

Finally we note that both  $\chi^1$ and  $\chi^2$ have a symmetric (under 
$\nu\leftrightarrow\mu$) and anti-symmetric terms.
For $\chi^1$ the symmetric part is on-shell and the anti-symmetric part is 
off-shell. The situation is opposite for $\chi^2$.
In general, the on-shell terms describe dissipation and vanish for gapped system 
and zero temperature while the off-shell terms describe the renormalization of 
the systems' parameters. For system with time-reversal symmetry all 
anti-symmetric contributions vanish \cite{FFprime} and, as a result, all odd 
susceptibilities are dissipative and on-shell while all the even 
susceptibilities are off-shell and describe renormalization of systems' 
parameters. These findings generalize the result of Berry and 
Robbins~\cite{berry_robbins}. Finally we note that all susceptibilities change 
sign with the temperature and that for positive temperature ($\beta>0$) the 
macroscopic degrees of freedom loses energy to the environment in all orders in 
the derivatives of $\vec{\lambda}(t)$.


\begin{thebibliography}{References}
\bibitem{mynote} Because we are taking the expectation value with the 
time-dependent density matrix $\rho_n(t)$ there will be corrections to the 
generalized force. However, one can show that these corrections are of the order 
of $|\dot{\vec\lambda}|^3$, which is beyond the accuracy considered in this 
paper.
\bibitem{sign} The minus signs are conventional. The sign of $F$ is chosen to 
reproduce known expressions for the Berry curvature.
\bibitem{FFprime} Using the Lehmann representation it is easy to write the 
antisymmetric tensors as differences of two-time correlation functions evaluated 
at $\pm t$. For system with time-reversal symmetry these correlation functions 
are identical and all antisymmetric tensors must vanish identically.
\bibitem{HObath} To obtain $\eta\ne0$ we need to assume that there are 
oscillators with vanishingly 
small frequencies $\omega_j\rightarrow 0$.
\end{thebibliography}
\end{document}